\documentclass[openacc]{rstransa} 
\usepackage[utf8]{inputenc}
\usepackage{amsmath, amssymb}
\usepackage{graphicx}
\usepackage{caption}
\usepackage{times}
\usepackage[numbers]{natbib}
\usepackage{orcidlink}

\usepackage{subcaption}
\titlehead{Research}
\begin{document}
\title{On the Modeling of Kink Oscillations in Fine-Structured Coronal Loops with Field-aligned Nonlinear Longitudinal Disturbances} 

\author{
Shakti Singh$^{1}$, 
Balveer Singh\orcidlink{0000-0001-6234-6400}$^{2}$, 
V. S. Pandey$^{1}$, 
Shilpa Patra$^{1}$,
Preeti Verma$^{1}$,
and Peter H. Yoon$^{3}$ }

\address{$^{1}$ Department of Physics, National Institute of Technology Delhi, New Delhi-110036, India\\
$^{2}$Division of Mathematics, School of Science and Engineering, University of Dundee, DD1 4HN, UK\\
$^{3}$Institute for Physical Science and Technology, University of Maryland College Park, MD, 20742, USA
}

\subject{Solar and Heliospherics}


\corres{Balveer Singh\\
\email{bsingh001@dundee.ac.uk singhbalveer37@gmail.com}}


\begin{abstract}
We investigate the influence of nonlinear longitudinal disturbances, triggered by an initial field-aligned velocity flow, on the damping of standing kink oscillations in fine-structured, cool coronal loop strands under isobaric conditions. Using two-dimensional ideal MHD simulations, we model different realistic flow geometries, bounded, unbounded, and external, and quantify their impact on wave excitation, damping, and energy leakage. Our modelled loop strand incorporates strong density contrasts ($d = 5$ and $d = 20$), providing a configuration consistent with cool-loop observations from Hinode/SOT and SDO/AIA. Our results show that while the nonlinear disturbances exert mild influence on the oscillation period, they substantially modify the damping time. Unbounded flows yield the strongest damping, reducing the damping time by up to $\sim 25\%$ compared to bounded flows, caused by enhanced wave-flow coupling and scattering. Longitudinally inhomogeneous flows further intensify the damping, with high-density strands exhibiting faster convergence toward the uniform-flow limit. In cases with supersonic internal flow, slow sausage-mode harmonics and weak slow shocks are additionally excited, indicating the generation of mixed-mode wave responses in flow-dominated loops. These findings demonstrate that realistic, spatially extended nonlinear disturbances play a significant role in the damping of kink oscillations and should be incorporated into forward modeling and coronal seismology diagnostics.
\end{abstract}

\maketitle
\section{Introduction}

Solar coronal loops are fundamental magnetic structures in the solar atmosphere, composed of hot, ionized plasma confined by magnetic fields. These loops exhibit arch-like geometries and occur across a wide range of spatial and temporal scales \citep{1999ApJ...520..880A, 2003ApJ...598.1375A}. These loops are dynamic sites of solar activity, hosting a variety of solar phenomena such as flares, coronal mass ejections (CMEs), and jet-like plasma flows. Coronal loops also support various magnetohydrodynamic (MHD) wave modes, oscillations, and instabilities, which play a crucial role in energy transport and heating in the solar corona \citep{2020PPCF...62a4016A, 2020ARA&A..58..441N}. Understanding the physical processes governing these loops is essential for explaining coronal heating and solar wind acceleration, which is the most enduring problems in solar physics. High-resolution observations from instruments such as the Atmospheric Imaging Assembly (AIA) onboard Solar Dynamics Observatory (SDO), TRACE, Hinode/SOT, and Solar Orbiter have revealed transverse oscillations in coronal loops, including both decaying and decayless modes \citep{2024MNRAS.531.4611N, 2024A&A...685A..36S}.These oscillations are interpreted as fast magnetosonic kink waves and provide main diagnostics for coronal seismology  \citep{2021SSRv..217...73N,2015ApJ...799..151G}. Their properties such as period, amplitude, and damping time can infer magnetic field strength, plasma density, and other key parameters of coronal loops \citep{2019ApJS..241...31N,2021SSRv..217...73N, 2008A&A...487L..17V}. Depending on their polarization, these oscillations can be classified as horizontal (perpendicular to the loop plane) or vertical (within the loop plane) \citep{1999Sci...285..862N, 2004A&A...421L..33W}. In particular, vertical kink oscillations have been observed in association with flare-induced disturbances and during the formation of post-flare loops, highlighting their connection with impulsive energy release processes\citep{2012A&A...545A.129W,2013ApJ...777...17S}. The ability to use these oscillations as diagnostic tools makes them central to era of high-resolution solar physics. 

High-resolution observations indicate the presence of field-aligned flows in coronal loops, which are generally sub-Alfv\'enic and less than about 10\% of the local Alfv\'en speed \citep[][]{2002ApJ...567L..89W, 2011Natur.475..477M}. These flows have been detected through Doppler shift measurements and feature tracking in imaging observations \citep{2001ApJ...553L..81W, 2008ApJ...689L..73C}. Such flows can affect the excitation, propagation, and damping of kink oscillations \citep{2015A&A...577A...4Z}. Proposed mechanisms for generating these flows include thermal instability leading to catastrophic cooling \citep{2003A&A...411..605M, 2010ApJ...716..154A} and pressure-driven siphon flows between loop footpoints \citep{2004A&A...427.1065T}. Recent observations further suggest that these flows are often structured and inhomogeneous, complicating their interaction with oscillations. This complexity highlights the need for models that incorporate realistic flow profiles and plasma conditions. The importance of flows in coronal loop dynamics extends to wave damping mechanisms such as resonant absorption and phase mixing \citep{2002ApJ...577..475R, 2002A&A...394L..39G, 2002ApJ...576L.153O}. Coronal seismology relies on these oscillations to infer magnetic field strength \citep{2001A&A...372L..53N}, density stratification \citep{2005ApJ...624L..57A}, and transverse density profiles \citep{2016A&A...589A.136P}. Accurate modeling of flows and oscillations is therefore critical for interpreting observational data and improving our understanding of coronal heating. Recent studies have shown that the presence of flows can significantly modify damping rates, yet most models assume simplified conditions that do not fully capture observed complexity. In particular, the role of flow geometry, bounded versus unbounded, remains poorly understood. 

Theoretical and numerical studies have explored the interaction of different flows with MHD waves. Earlier, TRACE observations confirmed the existence of kink oscillations \citep{1999ApJ...520..880A, 1999Sci...285..862N}, prompting extensive modeling efforts. Early slab models examined steady flows and their effect on kink and sausage modes \citep{1995SoPh..159..213N}, later extended to cylindrical geometries \citep{2003SoPh..217..199T}. These studies revealed that high-speed flows can trigger instabilities such as Kelvin–Helmholtz \citep{2008ApJ...687L.115T}. More recent work has addressed resonant damping, nonlinear mode coupling, and the influence of plasma-$\beta$, twist, and curvature on wave properties \citep{2015A&A...580A..57R, 2020MNRAS.496...67B, 2024ApJ...972...38L}. However, the combined effect of field-aligned flows and transverse inhomogeneity remains an open question, particularly under realistic coronal conditions. Furthermore, most previous studies have focused on strands with coronal temperature, whereas observations suggest that isobaric conditions may be more representative of coronal loops. Previous studies have largely ignored isobaric conditions with unbounded and external flows, limiting their applicability to real cool coronal loops. This study addresses these gaps by incorporating isobaric strands and multiple flow geometries, including unbounded and external flow profiles, which have not been systematically explored before.

Motivated by these gaps, the present study focuses on isobaric strands, which provide a more realistic representation of observed coronal loop conditions compared to traditional isothermal coronal models. Isobaric assumptions are supported by observational evidence indicating that pressure variations along many coronal loops are relatively small, making this approach physically consistent with solar atmospheric conditions. In this work, we investigate how different flow configurations—bounded, unbounded, and external- influence the damping of impulsively excited standing kink waves. These configurations are chosen to capture a range of plausible scenarios for plasma motion within and around the coronal loop's strand. Using two-dimensional numerical simulations that incorporate flow magnitudes consistent with observations, we analyze the excitation and damping of kink oscillations under realistic coronal conditions. Particular attention is given to the geometry of the initial flow and the resulting nonlinear disturbances, and their interaction with transverse wave dynamics, as these factors can significantly modify damping rates and energy transport. Our results provide new insights into how the structured nonlinear disturbances affect wave dissipation and highlight their implications for coronal seismology, especially in the context of diagnosing plasma parameters and magnetic field properties.

The present paper is organized as follows: Section~\ref{sec:model} describes the numerical model and simulation setup, Section~\ref{sec:Result} presents the main findings, and Section~\ref{sec:D&C} summarizes the conclusions and discusses future directions.

\section{Model and Methods} \label{sec:model}

\subsection{The MHD Model}
We consider a single-fluid MHD framework to model the cool coronal loop system in the solar atmosphere. The present model is gravity-free and accounts for a cool slab embedded in a solar chromospheric-like layer. The governing equations of the MHD system are given as follows:

\begin{equation}
    \frac{\partial \varrho}{\partial t} + \nabla \cdot \left( \varrho \mathbf{V} \right)=0, 
        \label{eq:mhd:1} 
\end{equation}

\begin{equation}
     \varrho\frac{\partial\textbf{V}}{\partial t} +\varrho(\textbf{V}\cdot\nabla)\textbf{V} = -\nabla p + \frac{1}{\mu}(\nabla\times\textbf{B})\times\textbf{B}, 
\end{equation}

\begin{equation}
  \frac{\partial\textbf{B}}{\partial t} = \nabla \times (\textbf{V} \times\textbf{B}),      \nabla \cdot \mathbf{B}=0,
\end{equation}

\begin{equation}
    \frac{\partial E}{\partial t} + \nabla \cdot \left[ \left( E + p_{T} \right){\bf V} - \frac{{\bf B}({\bf B} \cdot {\bf V})}{\mu} \right] = 0, 
        \label{eq:mhd:5} 
\end{equation}

\begin{equation}
    p = \frac{k_{\mathrm{B}}}{m} \varrho T 
        \label{eq:mhd:6}
\end{equation}

Here, '$\rho$' refers to the mass density. '$\mathbf{V}$' refers to the ion velocity, while '$\mathbf{B}$' represents the magnetic field. '$p$' represents the thermal pressure, and '$T$' is the temperature. '$k_{\mathrm{B}}$' is the Boltzmann constant, and '$m$' is the mean particle mass, fixed as $m = 1.24$. The symbol '$p_{t}$' is the total pressure, i.e., the sum of plasma pressure and magnetic pressure. The symbol '$E$' represents the total energy density of the system, which is given as follows. 
\begin{equation}
E = {\frac{p}{\gamma - 1}} + {\frac{\varrho v^2}{2}} + {\frac{B^2}{2 \mu}}.
\end{equation}
where the symbols '$\mu$' represent the magnetic permeability, '$\gamma$' is the ratio of specific heats given as $5/3$.

We numerically solved the above set of MHD equations using the static, grid-based ATHENA code \citep{2005JCoPh.205..509G}. In our implementation, the numerical fluxes were computed via a linearized Riemann solver, while the divergence-free condition for the magnetic field was maintained using the constrained transport (CT) method. The computational domain was defined as an Eulerian box with dimensions in the $x$--$z$ plane given by $(0,\,0.91\,L) \times (-0.5\,L,\,0.5\,L)$. This domain was discretized into a uniform grid comprising $300 \times 400$ cells.  We notice that our results are not sensitive to numerical resolution by performing initial tests with different grid sizes, which showed consistent oscillation characteristics. This behavior is in agreement with previous gravity-free slab models that demonstrated resolution-independent results \citep{2005A&A...440..385S, 2006A&A...454..653S, 2008A&A...489..413G}. Therefore, the chosen numerical setup does not significantly affect the robustness of our conclusions. Open boundary conditions were imposed on all boundaries, allowing outgoing wave signals to propagate freely out of the simulation region without reflection.

\begin{figure*}
\centering
\includegraphics[width=4.0cm, height=3.5cm]{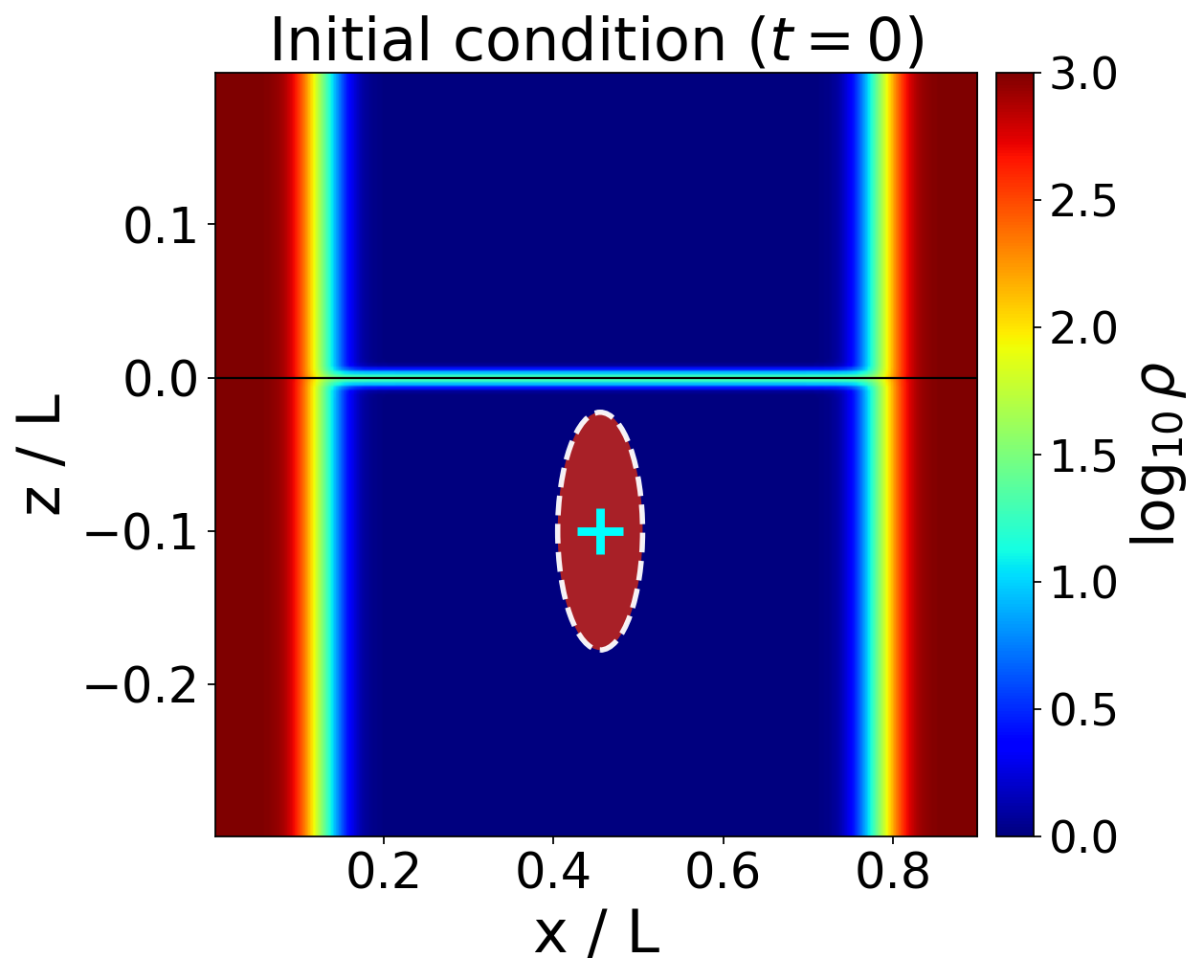}
\includegraphics[width=4.0cm, height=3.5cm]{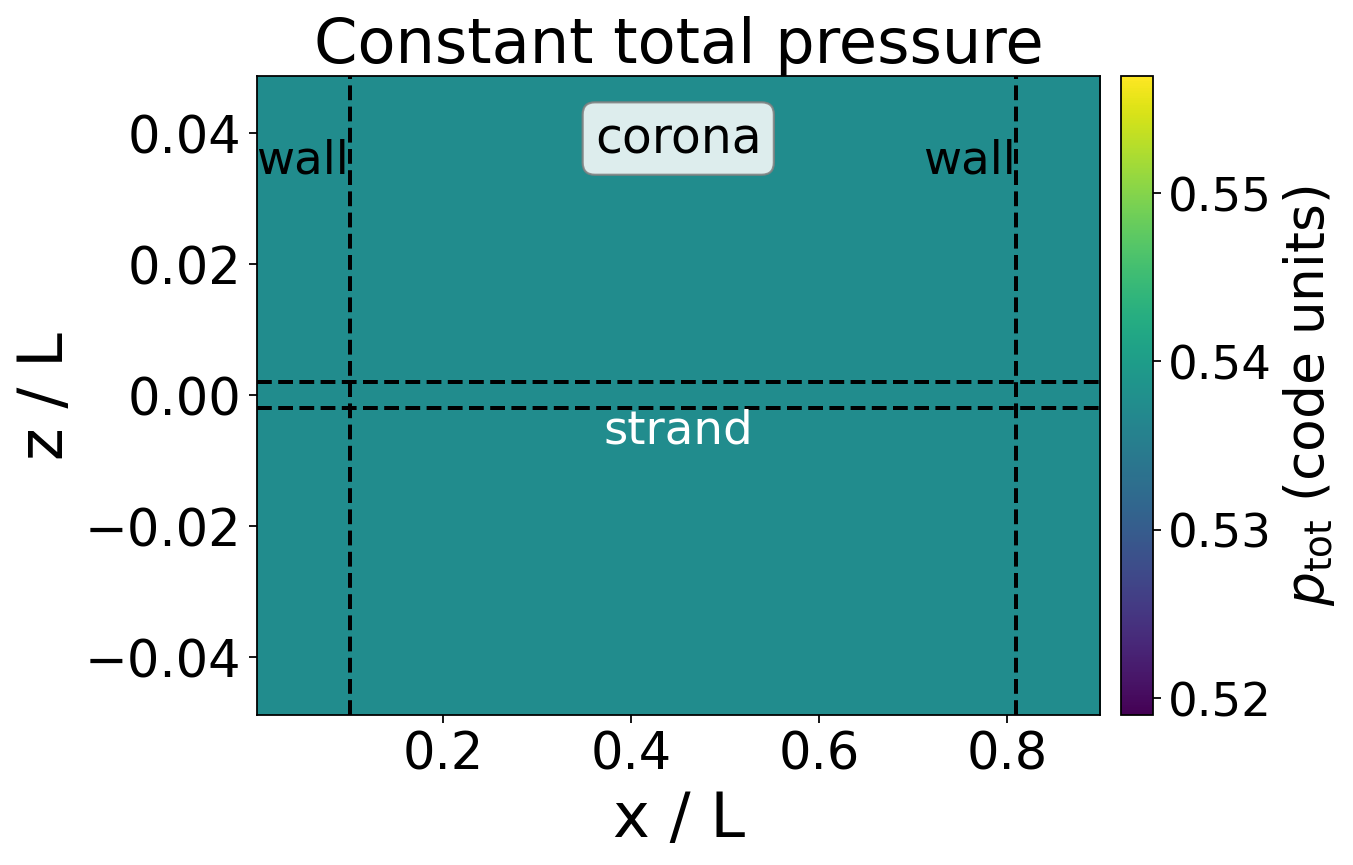}
\includegraphics[width=4.0cm, height=3.5cm]{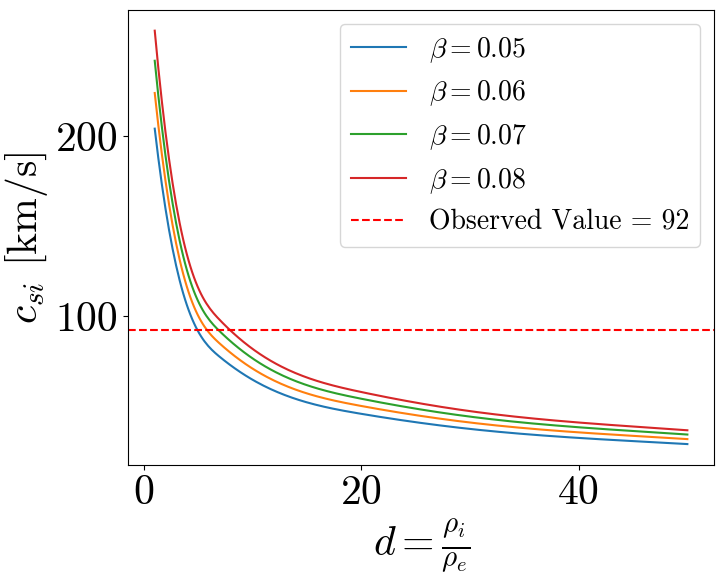}
\caption{Left: Logarithmic initial mass density (kg m$^{-3}$) profile of a straight coronal strand ($a = 0.4$ Mm) embedded in a loop of length $L = 100$ Mm, with dense chromospheric layers at $x/L < 0.10$ and $x/L > 0.81$. The dashed contour mark shows the location of the applied pulse at $z/L = -0.10$. Middle: Total pressure ($p_{\mathrm{gas}} + B^{2}/2\mu$ in normalized (code unit) at $t = 0$ s in the $x$--$z$ plane. The total pressure is uniform across the entire domain, the strand, the surrounding corona, and the dense chromospheric walls, confirming that the initial state is in exact total-pressure balance. The horizontal dashed black lines mark the strand boundaries ($z/L = \pm 0.002$, i.e.\ $z = \pm 0.2$~Mm) and the vertical dashed lines mark the chromospheric walls ($x/L = 0.10$ and $0.81$). For clarity, only the central region ($z/L \in [-0.05,\,0.05]$) is shown; the full domain extends to $z/L = \pm 0.5$. Right: Variation of the initial sound speed inside the strand as a function of density contrast of the strand for different values of ambient plasma $\beta$. The observed flow speed (92 km s$^{-1}$) from Hinode/SOT observations is marked by a horizontal red dashed line.}
\label{fig:fig1}
\end{figure*}

\begin{figure*}
\centering
\includegraphics[width=14.0cm, height=4.0cm]{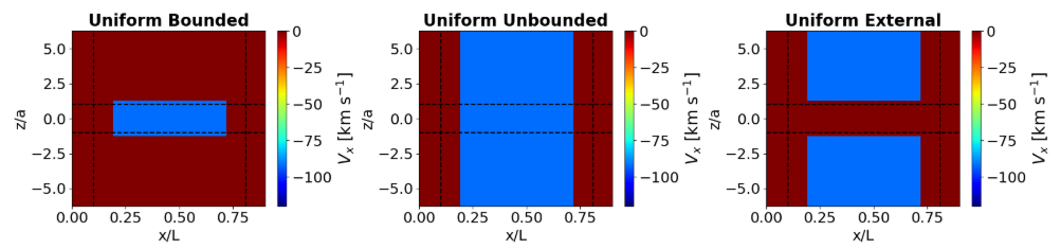}
\caption{Initial spatial distribution of the longitudinal velocity $V_{x}$ (km~s$^{-1}$) for the three uniform-flow configurations considered in this work, shown in the $x$--$z$ plane. Left: \emph{Uniform Bounded} flow, where $V_{x} = -92$~km~s$^{-1}$ is confined to the interior of the strand. Middle: \emph{Uniform Unbounded} flow, where the same magnitude of $V_{x}$ extends both inside the strand and throughout the surrounding ambient corona. Right: \emph{Uniform External} flow, where $V_{x}$ is present in the ambient corona but is excluded from the strand interior. The horizontal dashed lines indicate the boundaries of the strand at $z/L = \pm 0.002$, i.e.\ $z = \pm 0.2$~Mm, while the vertical dashed lines mark the locations of the dense chromospheric layers at $x = 0.10\,L$ and $x = 0.81\,L$.}
\label{fig:fig5}
\end{figure*}


\begin{figure*}[h]
\centering
\includegraphics[width=13.cm,height=8.0cm]{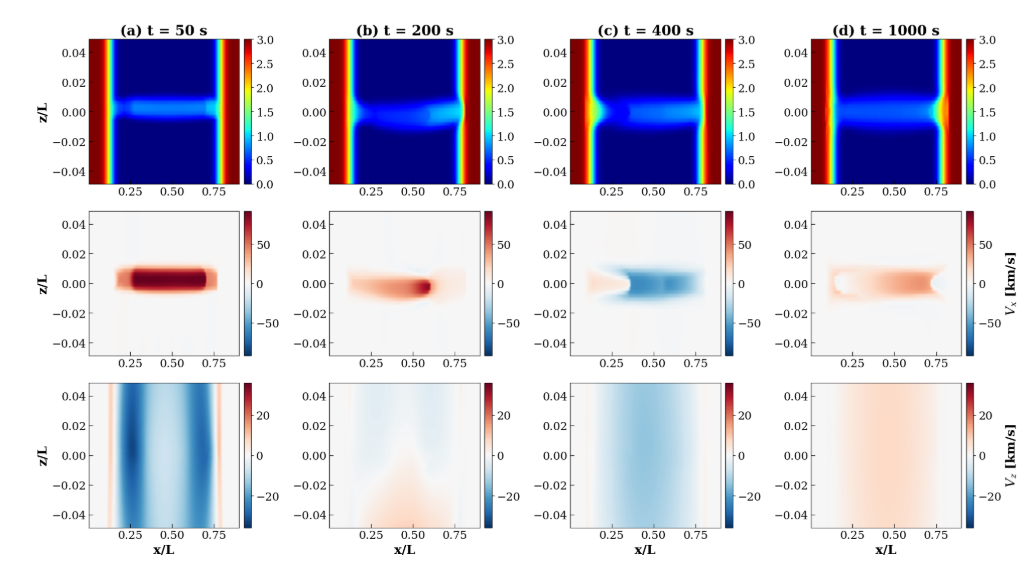}
\caption{Spatio-temporal evolution of plasma parameters within the coronal strand for the bounded uniform flow case at density contract d=5. Columns correspond to snapshots at (a) $t=50$ s, (b) $t=200$ s, (c) $t=400$ s, and (d) $t=1000$ s. The top row shows logarithmic mass density (kg m$^{-3}$), the middle row shows longitudinal velocity $V_x$ (km s$^{-1}$). The evolution illustrates transverse kink oscillations, progressive deformation of the strand, excitation of longitudinal disturbances, and subsequent damping of oscillatory motions. The reversal of $V_x$ indicates nonlinear evolution of the initial flow into field-aligned perturbations in the absence of external forcing. The bottom row presents the transverse velocity V${_z}$ exhibit spatially extended patterns that are not confined to the strand but spread across the surrounding plasma, indicating the global fast magnetoacoustic response to the impulsive perturbation. This global velocity field coexists with the localized transverse displacement of the density-enhanced strand, which constitutes the kink mode. For clarity, only the central region ($z/L \in [-0.05,\,0.05]$) is shown; the full domain extends to $z/L = \pm 0.5$.}

\label{fig:fig2}
\end{figure*}

\begin{figure*}[h]
\includegraphics[width=13.cm,height=6.0cm]{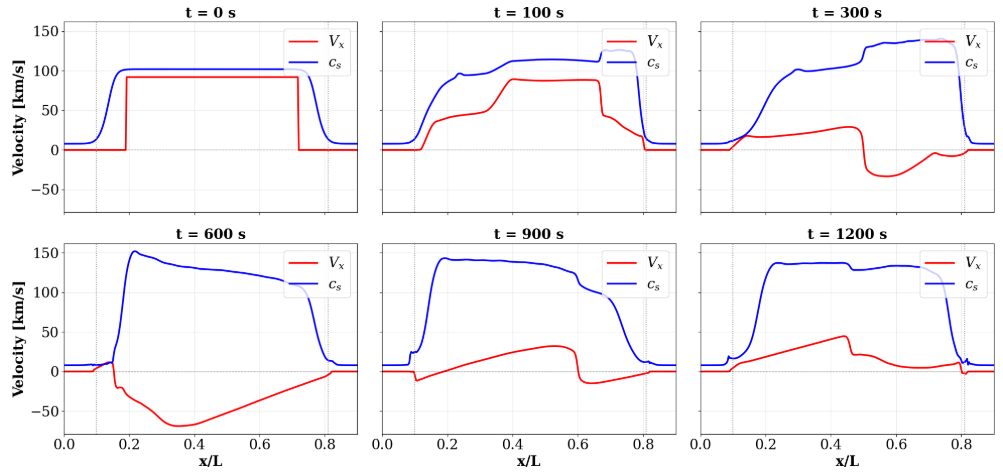}
\caption{Spatial evolution of the longitudinal velocity $V_x$ (red curve) and the local sound speed $c_s$ (blue curve) inside and around the coronal strand at six successive times: $t = 0$, $100$, $300$, $600$, $900$, and $1200$ s for the case of bounded uniform flow with density contrast, d=5. The flow is initialized with a positive $V_x$ inside the strand and is then allowed to evolve freely without any imposed boundary forcing. The longitudinal velocity progressively decreases, reverses sign, and redistributes itself along the loop, showing that the initial flow magnitude and direction are not preserved during the simulation. }
\label{fig:fig3a}
\end{figure*}

\subsection{Initial Configuration}
\label{sec:initial}

We adopt a two-dimensional slab model in the $x$--$z$ plane to represent a coronal loop strand of width $a = 0.4$~Mm. The strand is embedded in a loop with footpoints anchored in a dense chromospheric-like layer (see Fig.~\ref{fig:fig1}, left panel). The magnetic field is step-like and aligned along $x$, such that $B_x = B_i$ inside the strand. The multi-threaded nature of coronal loops indicates that individual strands oscillate collectively within the loop structure. Modeling flows confined to a single strand does not adequately represent this behavior. Therefore, we incorporate both internal and external field-aligned initial velocity perturbations around the fixed strand, whose nonlinear evolution into longitudinal disturbances influences the kink wave characteristics.

The strength of the magnetic field inside the strand is defined on the basis of conservation of total pressure (plasma + magnetic) across the boundary of the strand by the following relation: 
\begin{equation}
\frac{B_{i}}{B_{e}} = \ \sqrt{1 - \beta\left( \frac{p_{i}}{p_{e}} - 1 \right)}\ 
\end{equation}

Here, $p_i = p_e$ corresponds to an isobaric strand, which implies $B_i = B_e$. Thus, the internal magnetic field strength is independent of the density contrast $d$. In this setup, both gas pressure and magnetic field remain uniform throughout the domain (see Fig.~\ref{fig:fig1}, middle panel), while the density alone introduces the contrast $d$. Consequently, the strand represents a pure density (and hence temperature, since $T \propto p/\rho$) enhancement under constant total pressure. The initial equilibrium therefore satisfies exact total pressure balance, $p_{\mathrm{gas}} + B^{2}/2\mu = \mathrm{const}$, as shown in Fig.~\ref{fig:fig1}. 
We choose the density contrast, $d=5$ and  $d=20$; under the isobaric condition this reduces the strand temperature by the same factor of  $d=5$ and  $20$ , respectively, relative to the surrounding coronal loop plasma. The observations made by  Ofman \& Wang 2008 \citep{2008A&A...482L...9O} and Antolin \& Verwichte 2011 \citep{2011ApJ...736..121A} correspond to chromospheric plasma (temperature \(\sim 10^{4}\)) K for the strand. It is very likely that the immediate surroundings of the cool plasma may be at transition region temperatures or even at coronal temperatures \citep{2011ApJ...732...81A, 2013A&A...556A.104P, 2019A&A...627A..62G}, hence a temperature difference of 10-100. The sound speed (\(c_{Se}\)) in the ambient coronal medium is taken to be fixed and the different choices of the reference sound speed of the ambient medium yield different values of sound speed inside the isobaric strand by the following relationship: 
\begin{equation}
c_{Si} = \frac{c_{Se}}{\sqrt{d}} = v_{Ae}\sqrt{\frac{\gamma\beta}{2d}},
\end{equation}
where $\beta = \frac{2}{\gamma}\left( \frac{c_{Se}^{2}}{v_{Ae}^{2}} \right)$ is the plasma beta parameter for the ambient medium. The smooth variation of the sound speed inside the isobaric strand as a function of density contrast of the strand at different values of ambient plasma \(\beta\), is as shown in Fig.~\ref{fig:fig1}, right-panel. It is evident that the values of the sound speed inside the strand decrease with the increase in the value of density contrast of the strand. The value of the density contrast for which the sound speed inside the strand becomes less than or equal to the observed flow speed (horizontal dashed line) is determined by the value of the plasma \(\beta\) of the ambient medium.

To mimic the cool coronal loop, we include a dense plasma layer following the configuration of  Gruszecki et al.(2008a)\citep{2008A&A...488..757G}. The layer extends between $x_l = 0.10L$ and $x_r = 0.81L$, where $L = 100$~Mm is the total system length. The equilibrium density distribution is given by
\begin{equation}
\rho(x,z) = \rho_e + \frac{1}{2}\rho_e(z)(d_p - 1) \Big[ 
\left(1 - \tanh\left(\frac{x - x_l}{s_p}\right)\right) 
+ \left(1 + \tanh\left(\frac{x - x_r}{s_p}\right)\right) 
\Big], \\
\label{eq:density_profile}
\end{equation}
where $d_p = 10^3$ is the chromosphere-to-corona density ratio and $s_p = 2$~Mm denotes the transition width. The ratio of the mass density of the chromosphere to the ambient medium (corona outside the strand) is taken as\(\ d_{p} = 10^{3}\) with\(\ s_{p}\)= 2.0 Mm denoting the width of the transition region. These choices are quite significant in the context of present consideration that relates to the observations because amplitude of the strand displacement depends on the value of \(d_{p}\). Also, it is noteworthy that the ambient medium is customarily
referred to as the diffuse corona. In this study, the ambient medium is referred to as describing the environment surrounding a strand, inside the loop hosting the strand. Gruszecki et al.(2008a) \citep{2008A&A...489..413G}, in their study have shown that inclusion of dense chromospheric-like layer has a significant impact on the characteristics of waves. Magnetic field strength of \(B_{e}\)=11.21 G associated with the ambient medium is connected with the reference mass density \(\rho_{e}\) through the reference Alfv\'en speed
\(v_{Ae} = \ \frac{B_{e}}{\sqrt{\mu\rho_{e}}}\), where we choose
\(\rho_{e}\) = \(10^{- 12}\) kg \({\rm m}^{- 3}\)and \(v_{Ae}\)= 1000 km \(s^{- 1}\) for the mass density and the Alfv\'en speed, respectively. 

We consider field-aligned flow profiles, expressed as
\begin{equation}
v_{x0}(x) = v_0 \exp\left[-\frac{(x - x_0)^2}{\omega_{fx}^2}\right]\hat{\mathbf{x}},
\label{eq:longitudinal_flow}
\end{equation}
where $\omega_{fx}$ denotes the half-width; $\omega_{fx} \rightarrow \infty$ yields a uniform flow. This longitudinal profile was previously adopted by Gruszecki et al.(2008a) \citep{2008A&A...488..757G}, where the flow was confined within the strand thickness. 

The field-aligned flow imposed in our model is an initial condition rather than a driven, steady state. Owing to the substantial initial velocity perturbation (of the order of the local sound speed), this flow does not remain steady; it evolves nonlinearly, launching longitudinal disturbances that propagate along the strand at speeds approaching the local sound speed. It is these nonlinear longitudinal disturbances, and their interaction with the transverse kink motion, that govern the damping.

\subsection{Perturbation}
In this work, we focus on impulsively excited kink waves. To initiate these waves, we impose an initial perturbation on the vertical velocity component, $V_{z}$, expressed as
\begin{equation}
\ V_{z} = A_{z0}\,\exp\left[-\frac{(x - x_{0})^{2} + (z - z_{0})^{2}}{\omega^{2}}\right],
\end{equation}

where the symbol '$\omega$' depicts the Gaussian width of the pulse, fixed as $10~\mathrm{Mm}$. The perturbation is centered at $(x_{0}, z_{0})$ with amplitude $A_{z0} = 0.15\,V_{A0}$, where $ V_{A0} $ represents the Alfv\'en speed. The applied amplitude ($A_{z0} = 0.15\,V_{A0}$), corresponding to $\sim$ 150 km s$^{-1}$. This is lower than the typical coronal sound speed ($\sim$ 200 km s$^{-1}$), indicating that the perturbation is subsonic. In our setup, the pulse is applied below the slab at $x_{0} = 0.455\,L$ and $z_{0} = -0.1\,L$. Such localized disturbances are representative of flare-induced pulses near loop footpoints. While the primary response consists of kink waves, a minor contribution from slow-mode waves is also present.
%

\begin{table*}
   \caption{Wave characteristics with uniform flow for an isobaric strand at density contrast ($d$), which is equal to 5 and 20. These parameters are obtained after fitting of the decaying kink oscillations by making use of Eq.~\ref{fitting_eqn}}
   \resizebox{1.0\textwidth}{!}{ 
    \begin{tabular}{lclcccc}
     \hline\hline
        & \textbf{Density contrast ($d$)} & \textbf{Type of Flows} & \textbf{$A_{\rm max}$} & \textbf{P} & \textbf{$\tau$} & \textbf{$\frac{\tau}{P}$} \\ 
        & & & \textbf{(Mm)} &  \textbf{(sec)} & \textbf{(s)} \\ 
\hline 
 \hline
        
           Isobaric & $d=5$ & {No flows} & {0.474} & {108.3} & {709.9} & {6.56} \\ 
           & & Bounded flows & 0.474 & 108.3 & 705.4 & 6.52 \\ 
           & & Unbounded flows & 0.472 & 109.3 & 571.8 & 5.23 \\ 
           & & Outside the strand & 0.472 & 109.2 & 575.8 & 5.27 \\ \hline
    
          Isobaric & $d=20$ & {No flows} & {0.478} & {116.4} & {780.8} & {6.71} \\ 
          & & Bounded flows & 0.479 & 115.1 & 710.5 & 6.17 \\ 
          & & Unbounded flows & 0.475 & 116.0 & 608.4 & 5.24 \\ 
          & & Outside the strand & 0.476 & 116.9 & 680.2 & 5.82 \\ \hline 

    \end{tabular}
    \label{tab:uniformflow}}
\end{table*}


\begin{figure*}[h]
\centering
\includegraphics[width=13.0cm,height=4.5cm]{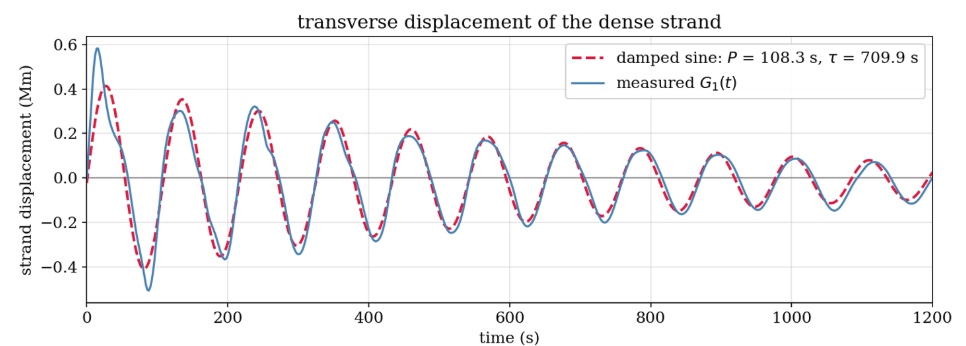}
\includegraphics[width=14.0cm,height=4.5cm]{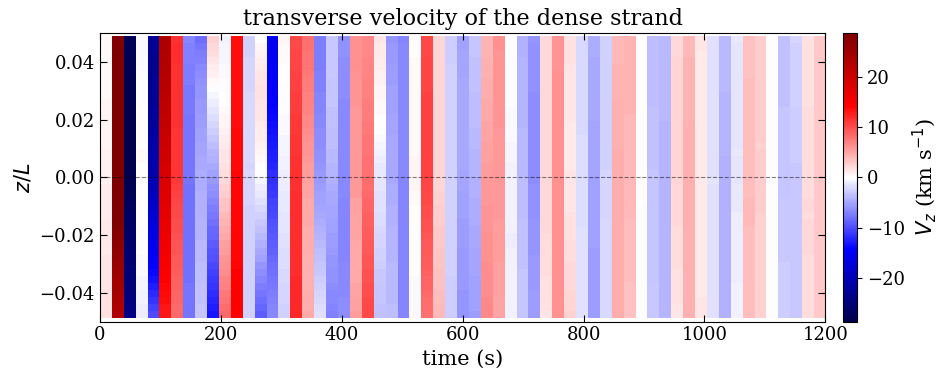}
\caption{Top: Transverse displacement of the dense strand for the bounded uniform flow case with $d=5$, obtained from the fitted Gaussian center $G_1(t)$ of Eq.~(\ref{eq:gaussian_fit}) (blue). The red dashed curve is the damped-sine fit of Eq.~(\ref{fitting_eqn}), yielding a period $P = 108.3$~s and damping time $\tau = 709.9$~s. The coherent damped oscillation of the strand axis demonstrates that the density-enhanced strand oscillates as a single entity. Bottom: Time–distance diagram of the transverse velocity component $V_z$ across the strand at the loop midpoint for the same case as the top panel. The alternating red–blue bands indicate periodic transverse motions extending across the computational domain, representing the global wave motion response to the initial impulsive perturbation.
} 
\label{fig:g1_vz}
\end{figure*}

\begin{figure*}
\centering

 \includegraphics[width=6.5cm,height=2.5cm]{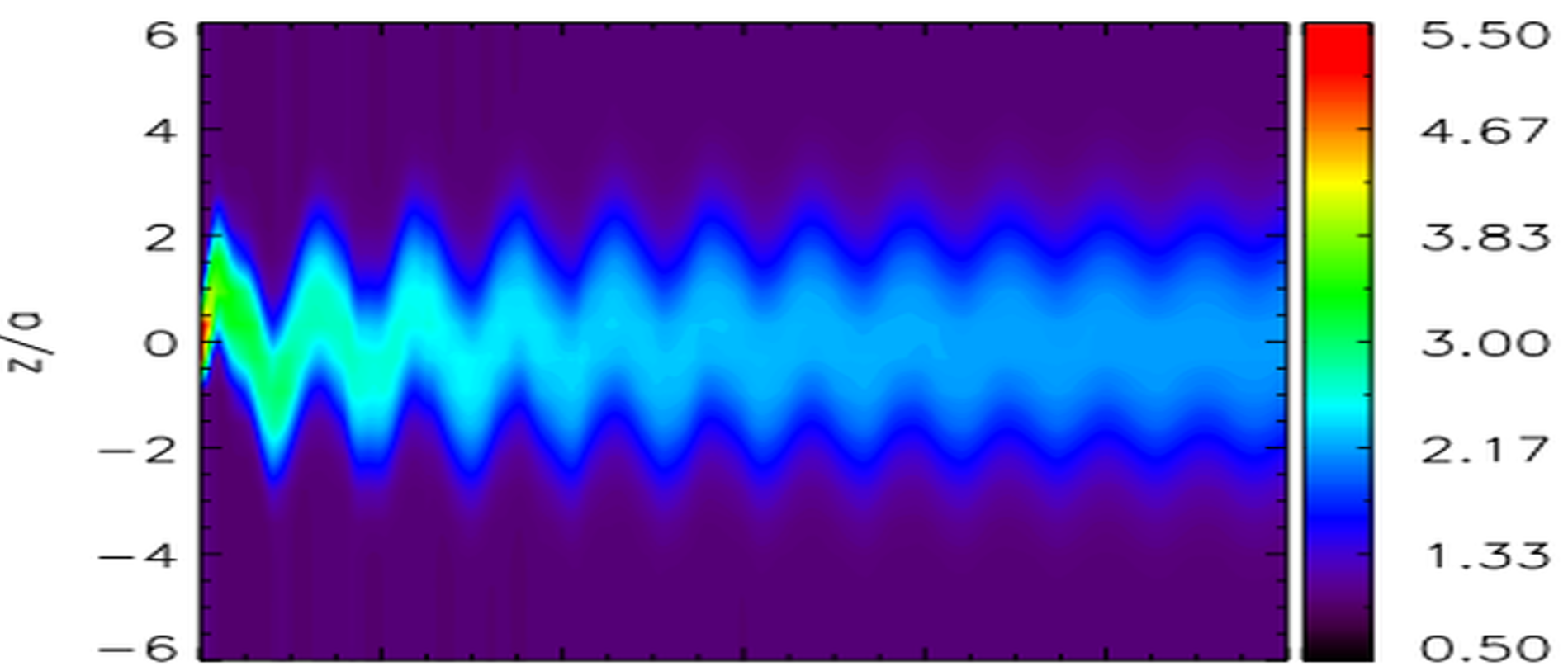}
 \includegraphics[width=6.5cm,height=2.5cm]{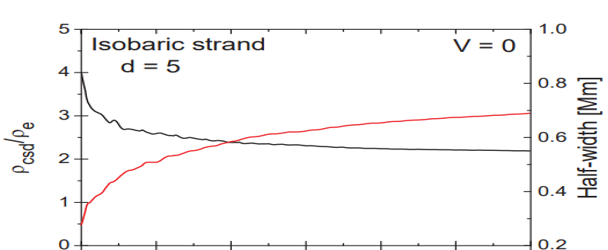}

\includegraphics[width=6.5cm,height=2.5cm]{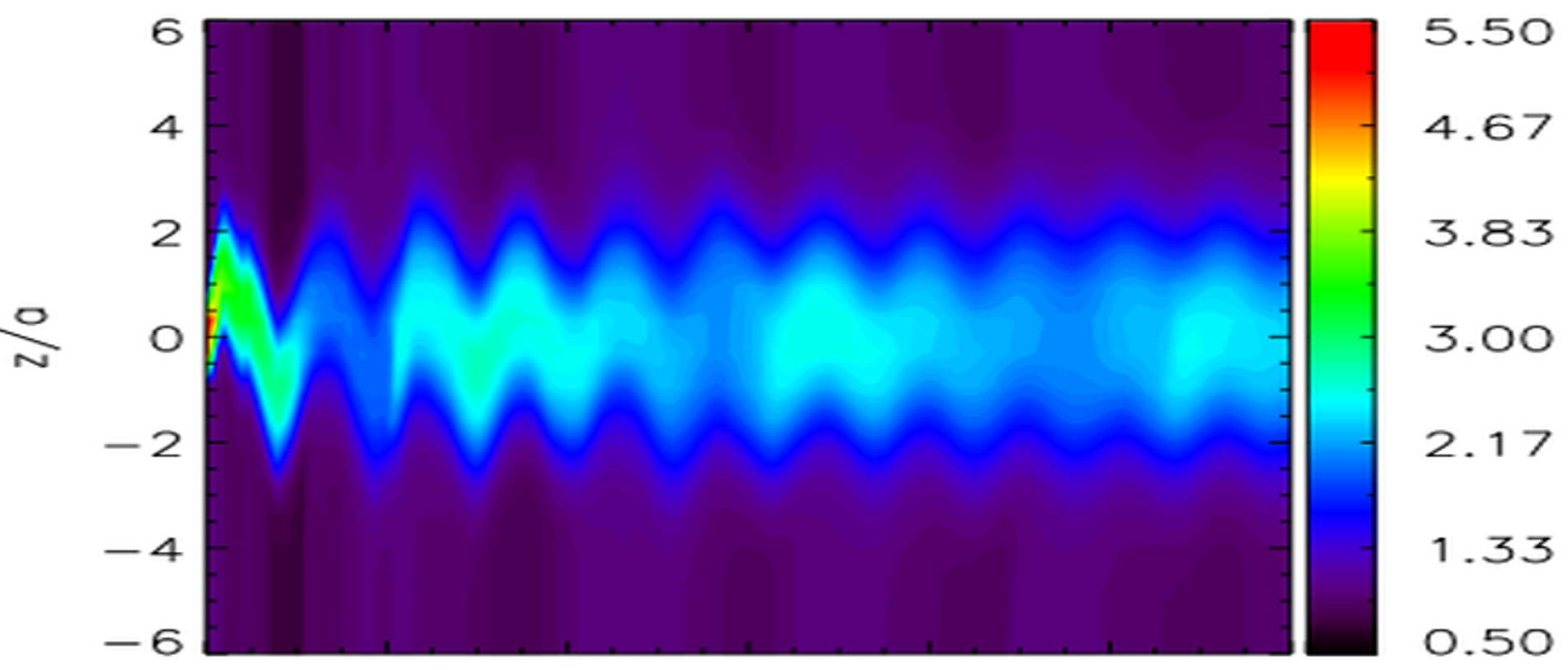}
\includegraphics[width=6.5cm,height=2.5cm]{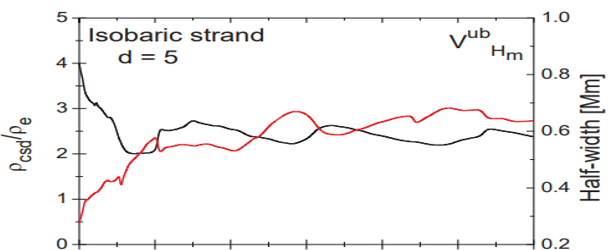}

\includegraphics[width=6.5cm,height=2.5cm]{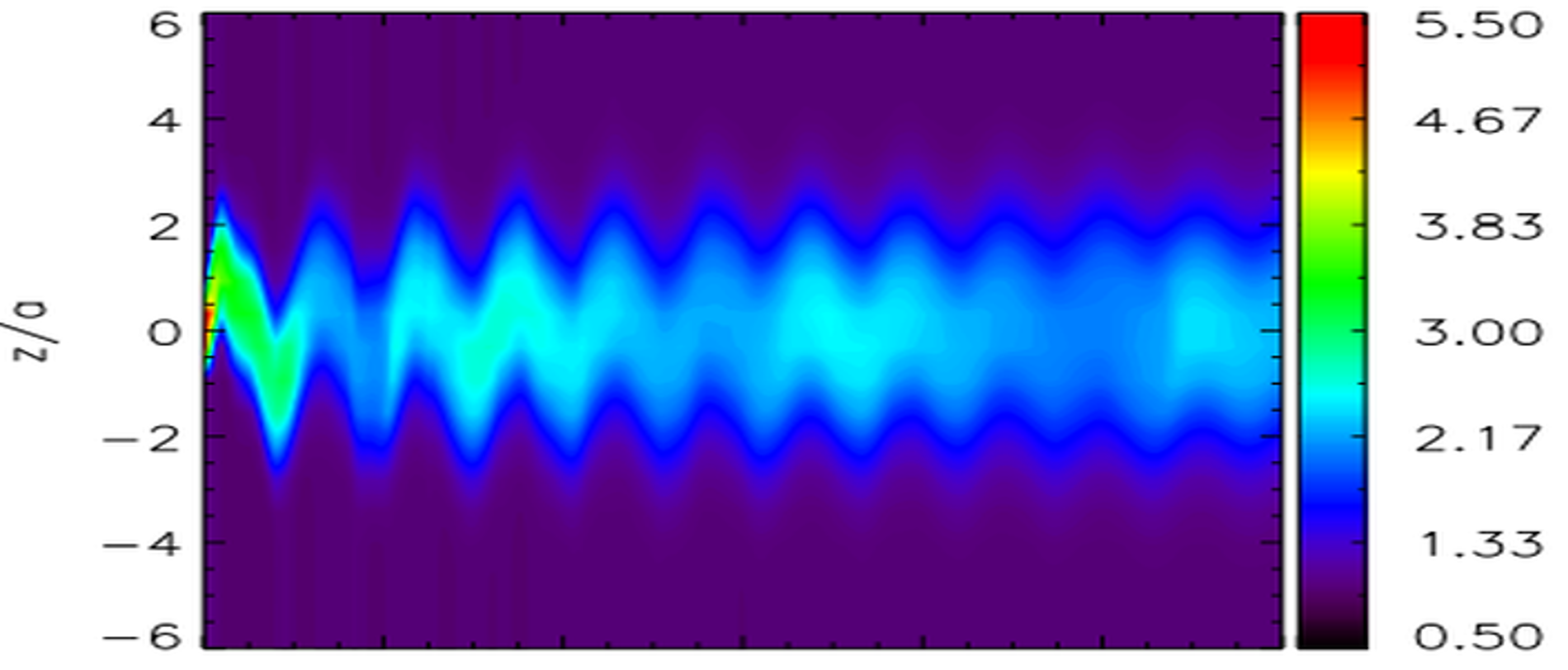}
\includegraphics[width=6.5cm,height=2.5cm]{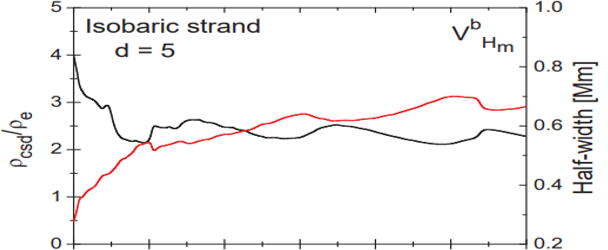}

\includegraphics[width=6.5cm,height=2.9cm]{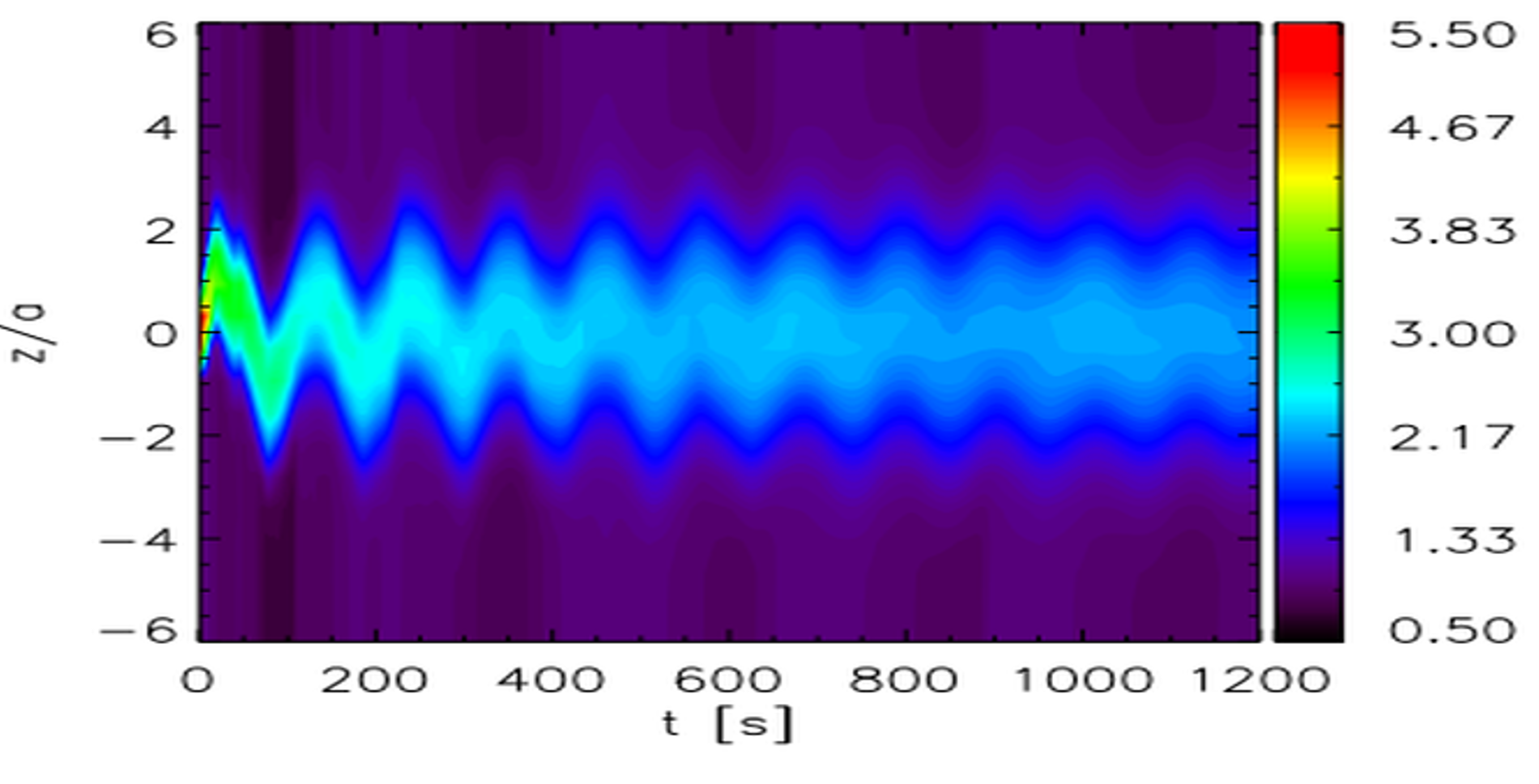}
\includegraphics[width=6.5cm,height=2.9cm]{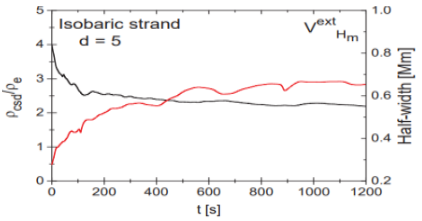}

\caption{Left-column: Time signatures of the mass density at the middle of the isobaric strand, where the density contrast of 5 with respect to the ambient coronal loop plasma of $\beta = 0.075$ is shown. Right-column: The temporal variation of the mass density of the strand (black curve on left $y$-axis) and half-width (red curve on right $y$-axis) at the middle of the isobaric strand. The first row panel corresponds to no flow (static case). The second and third rows correspond to a uniform initial flow of 92 km s$^{-1}$, localized either within the width of the strand or both inside and outside the strand. The bottom row corresponds to the uniform flow of the same magnitude pervading only outside the strand but not existing inside the strand.}
\label{fig:density_5}
\end{figure*}

\begin{figure*}
\centering
\includegraphics[width=6.5cm,height=2.5cm]{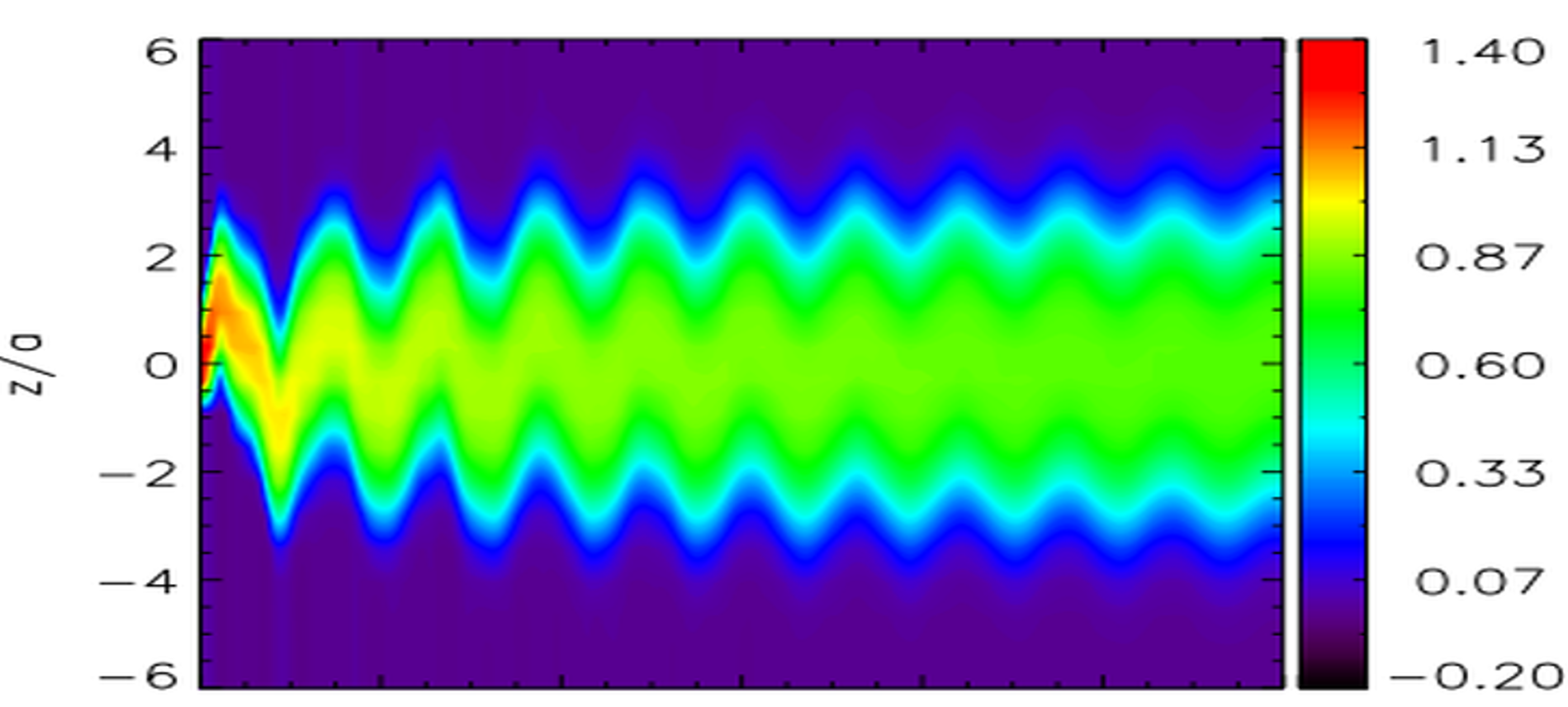}
\includegraphics[width=6.5cm,height=2.5cm]{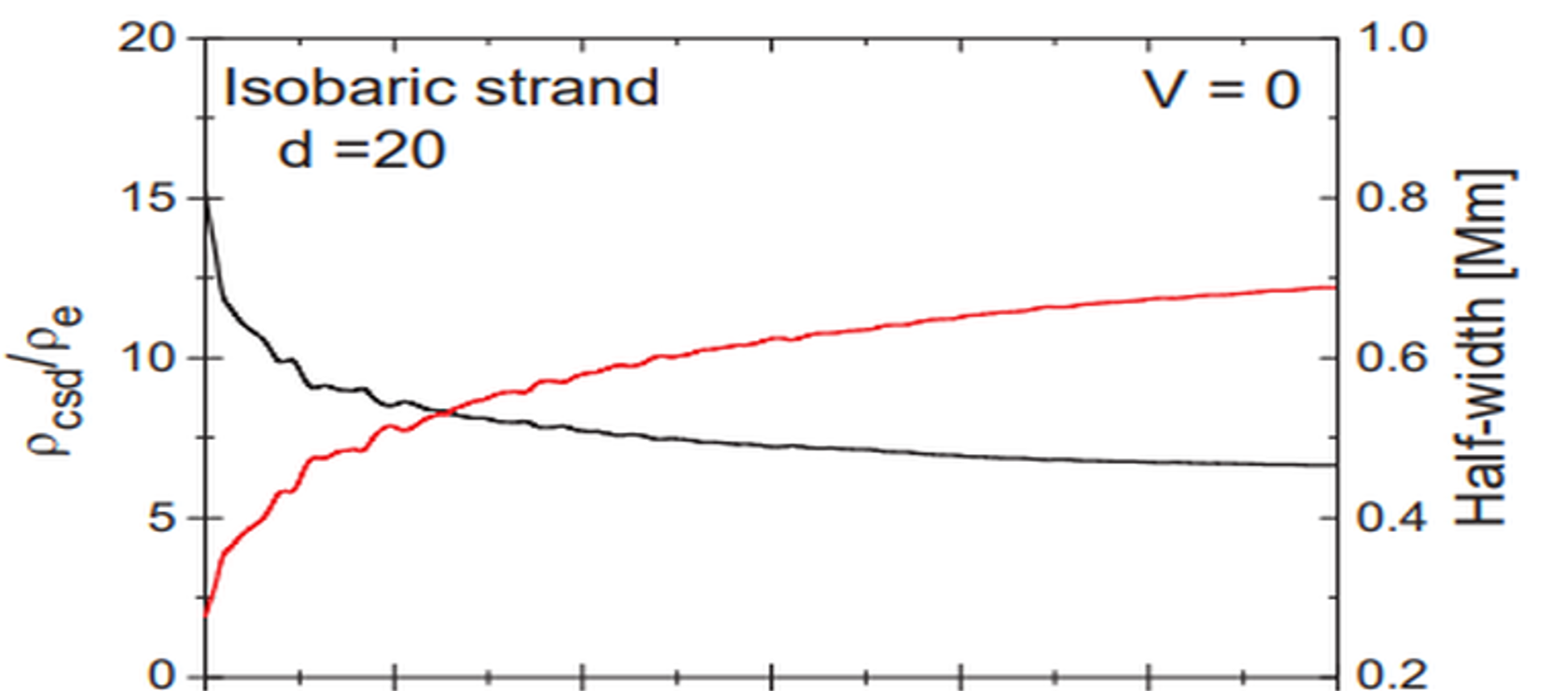}

\includegraphics[width=6.5cm,height=2.5cm]{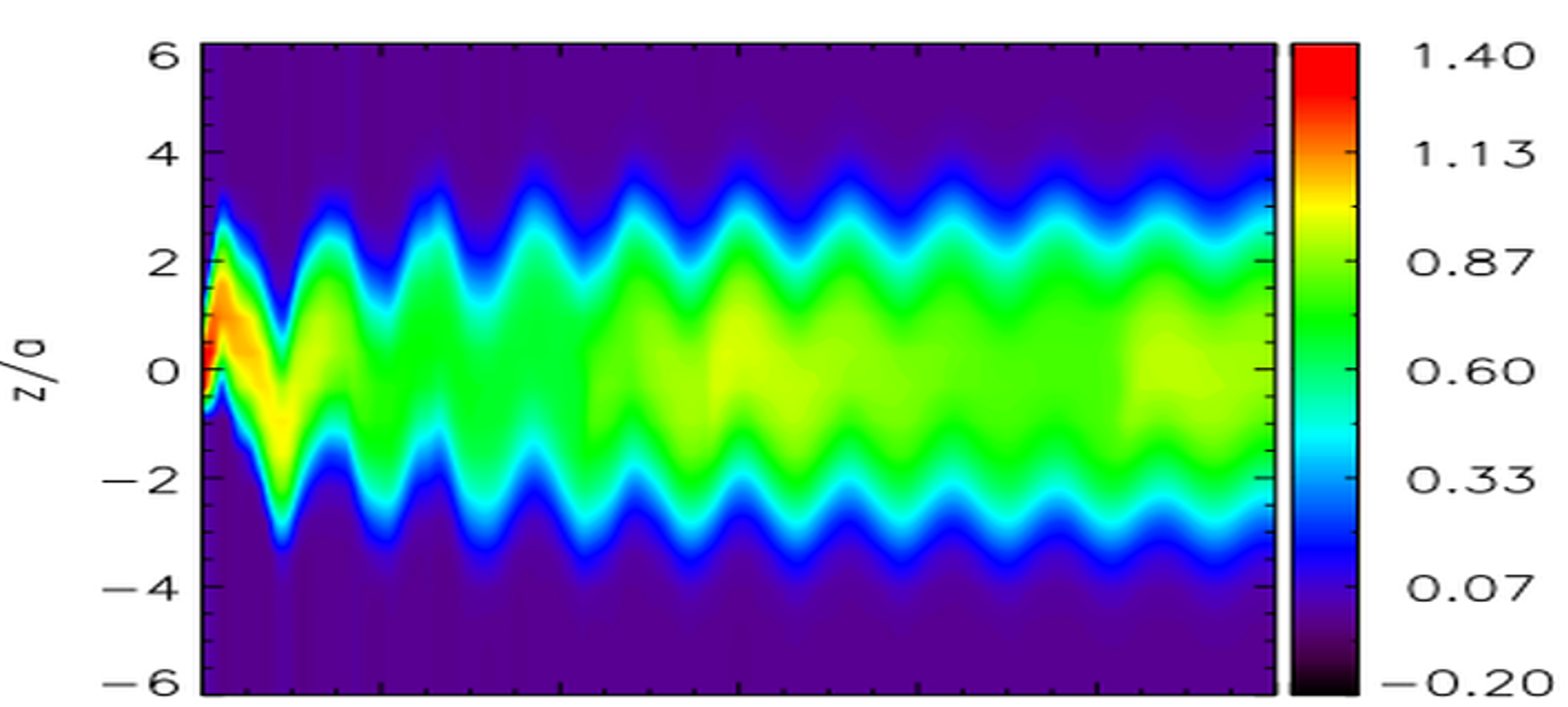}
\includegraphics[width=6.5cm,height=2.5cm]{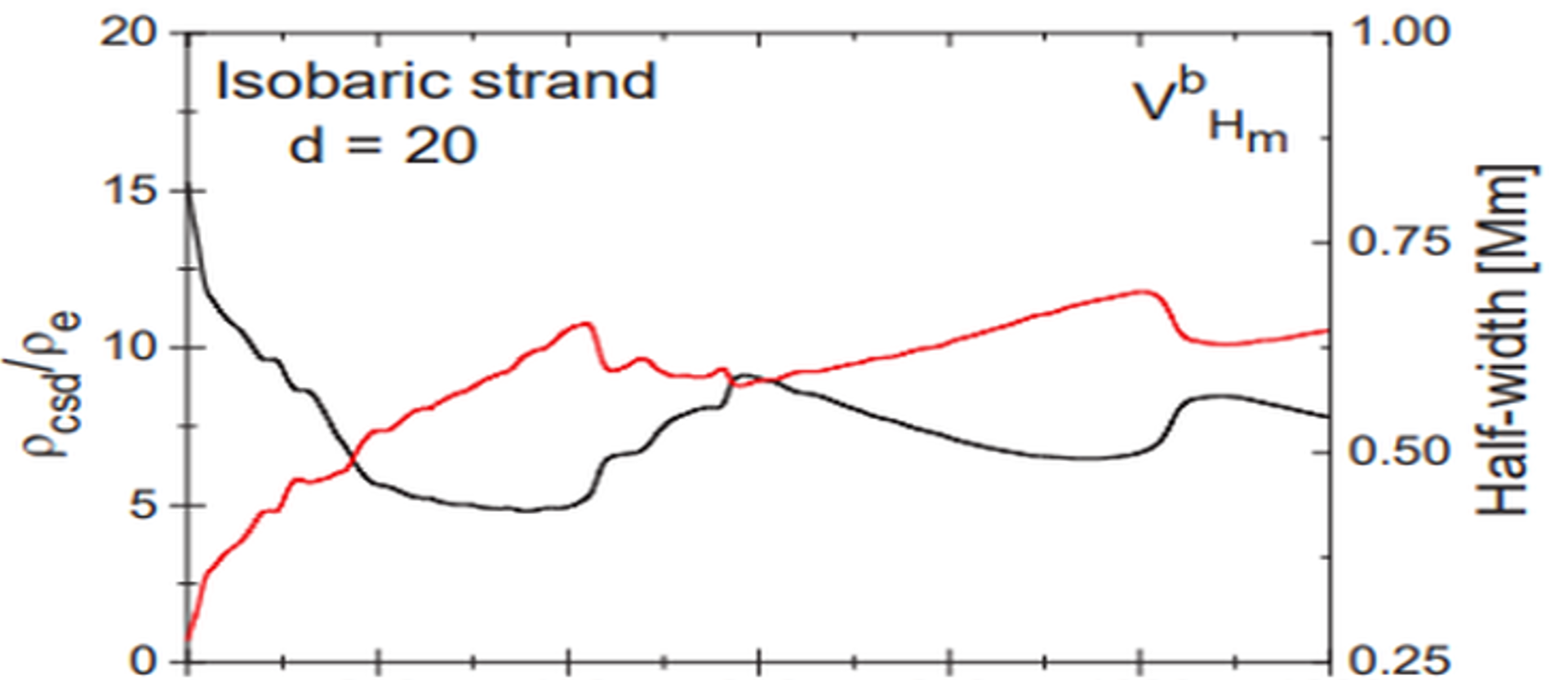}

\includegraphics[width=6.5cm,height=2.5cm]{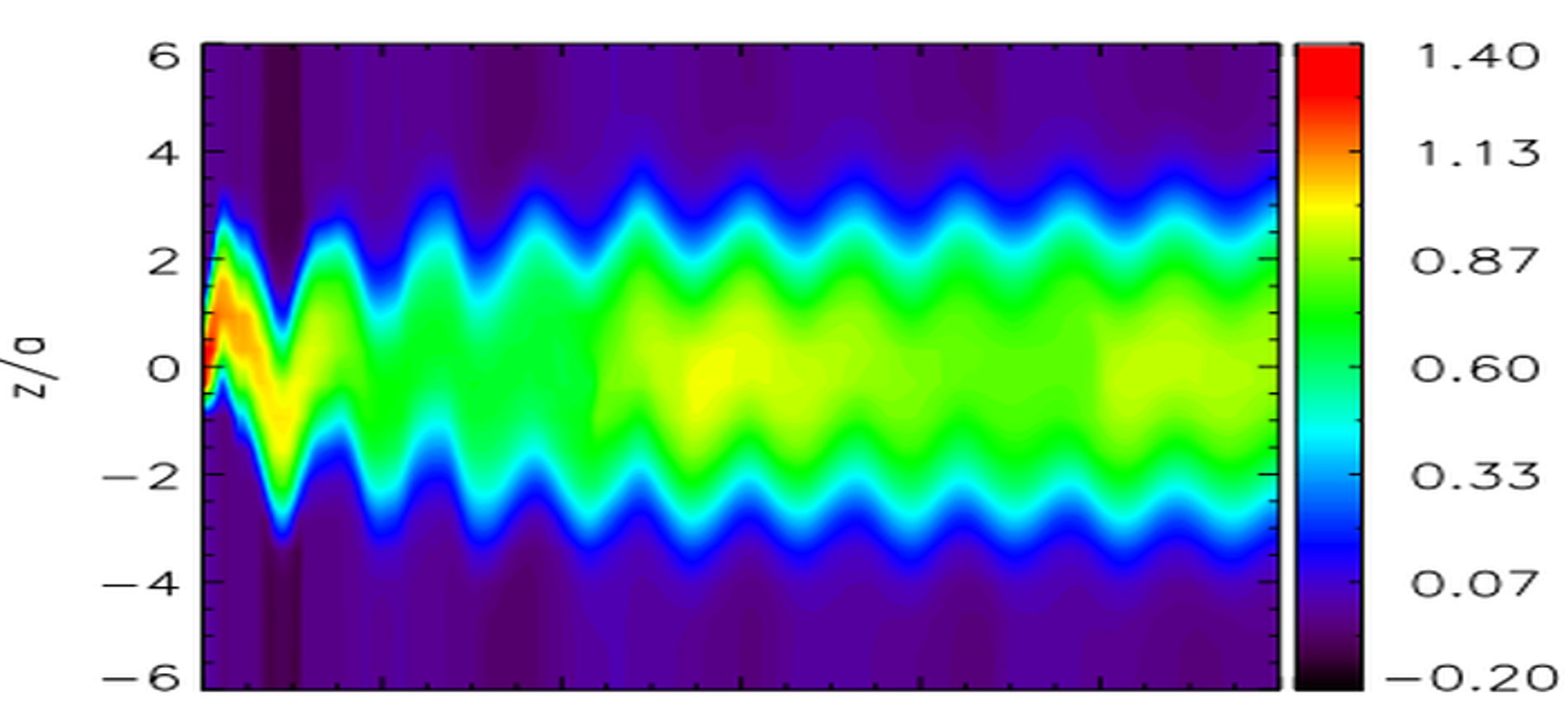}
\includegraphics[width=6.5cm,height=2.5cm]{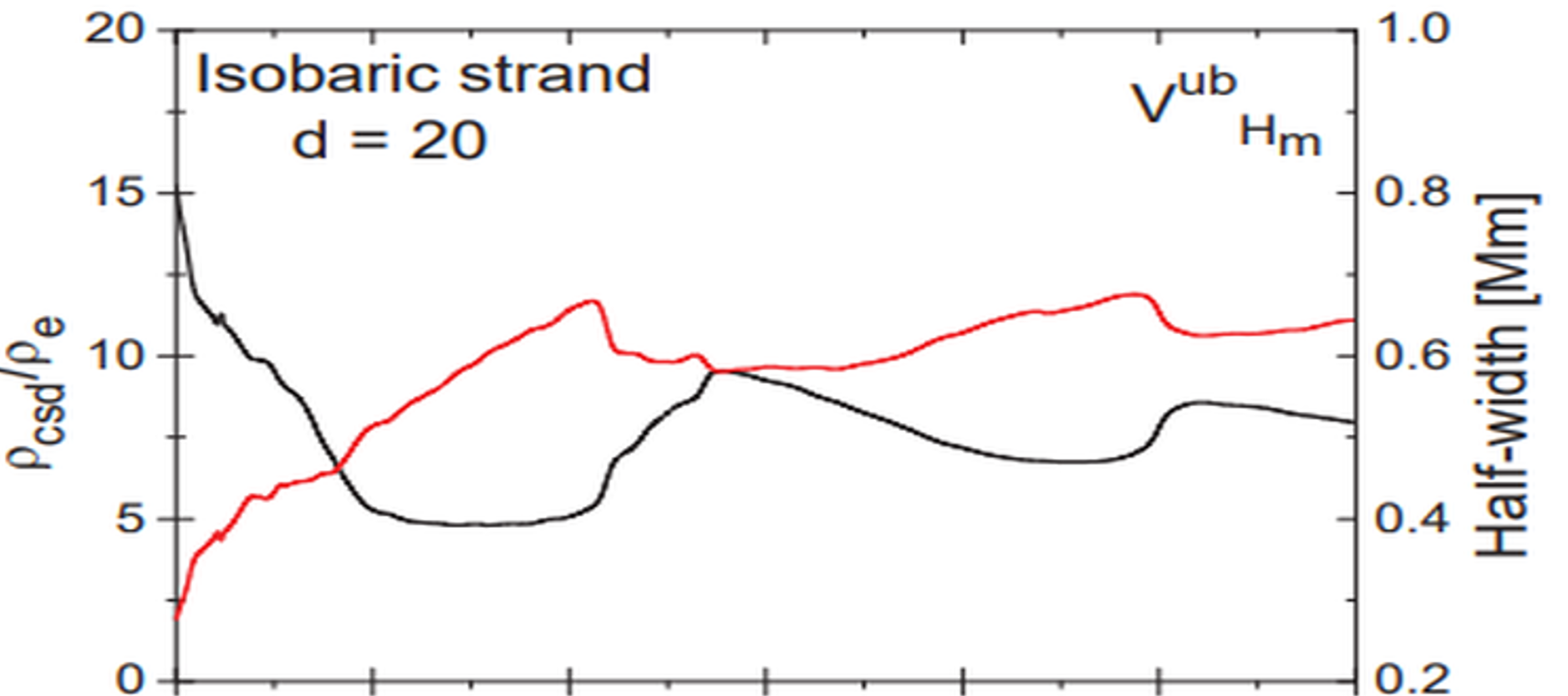}

\includegraphics[width=6.5cm,height=2.9cm]{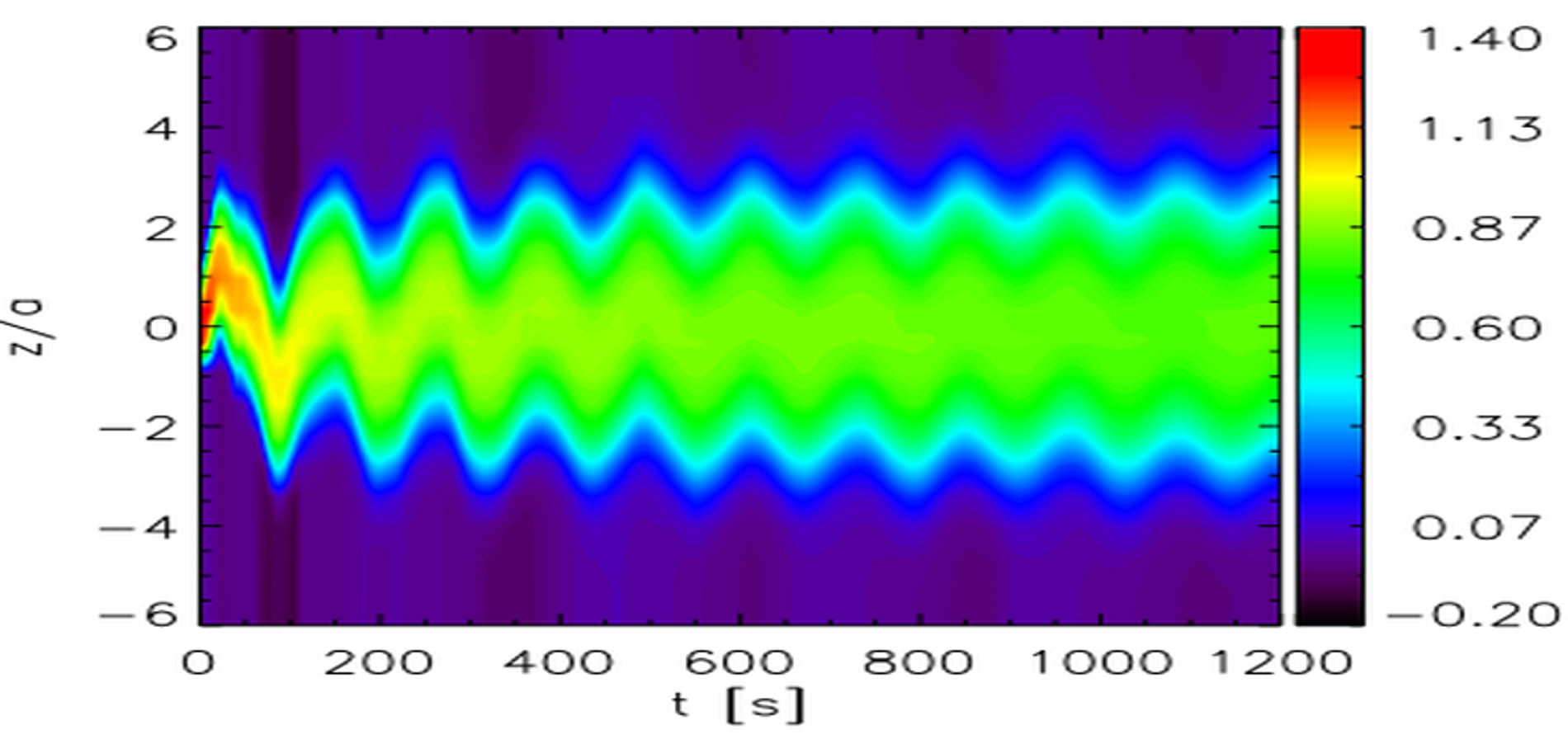}
\includegraphics[width=6.5cm,height=2.9cm]{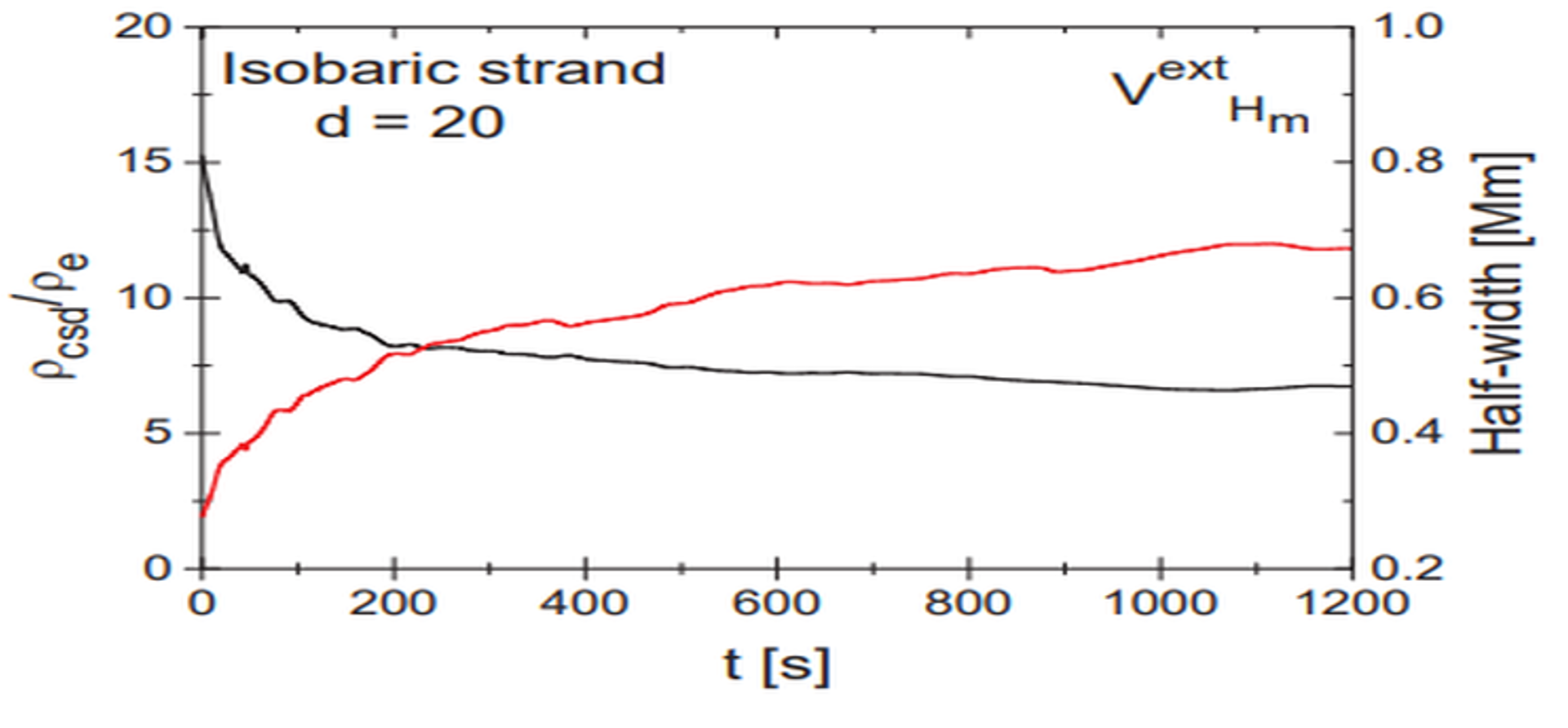}

\caption{Same as shown in Fig.~\ref{fig:density_5}, except that the density contrast of the strand is equal to 20. The stronger density contrast enhances slow‑mode signatures, including weak slow‑shock behaviour under supersonic internal flow.}
\label{fig:density_contrast20}
\end{figure*}

\begin{figure*}
\centering
\includegraphics[width=6.5cm,height=4.0cm]{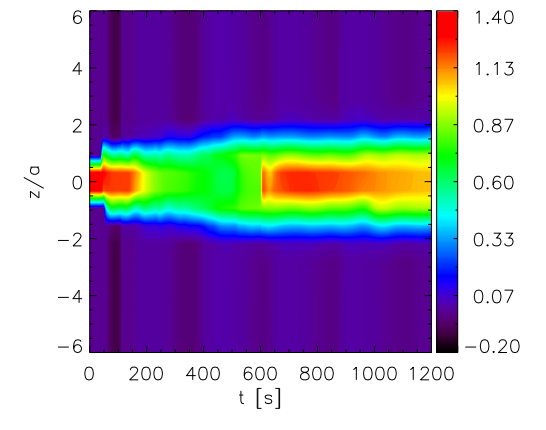}
\includegraphics[width=6.5cm,height=4.0cm]{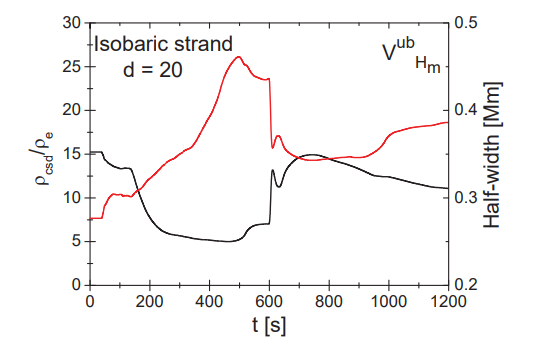}
\caption{Temporal evolution of mass density (left) and half‑width at the strand (right) center for the case of unbounded flow without an initial pulse, shown for density contrast d=20. The inherent flow alone excites slow sausage‑mode harmonics.}
\label{fig:unbounded_nopulse}
\end{figure*}

\begin{figure*}
\centering
\includegraphics[width=13.0cm, height=4.0cm]{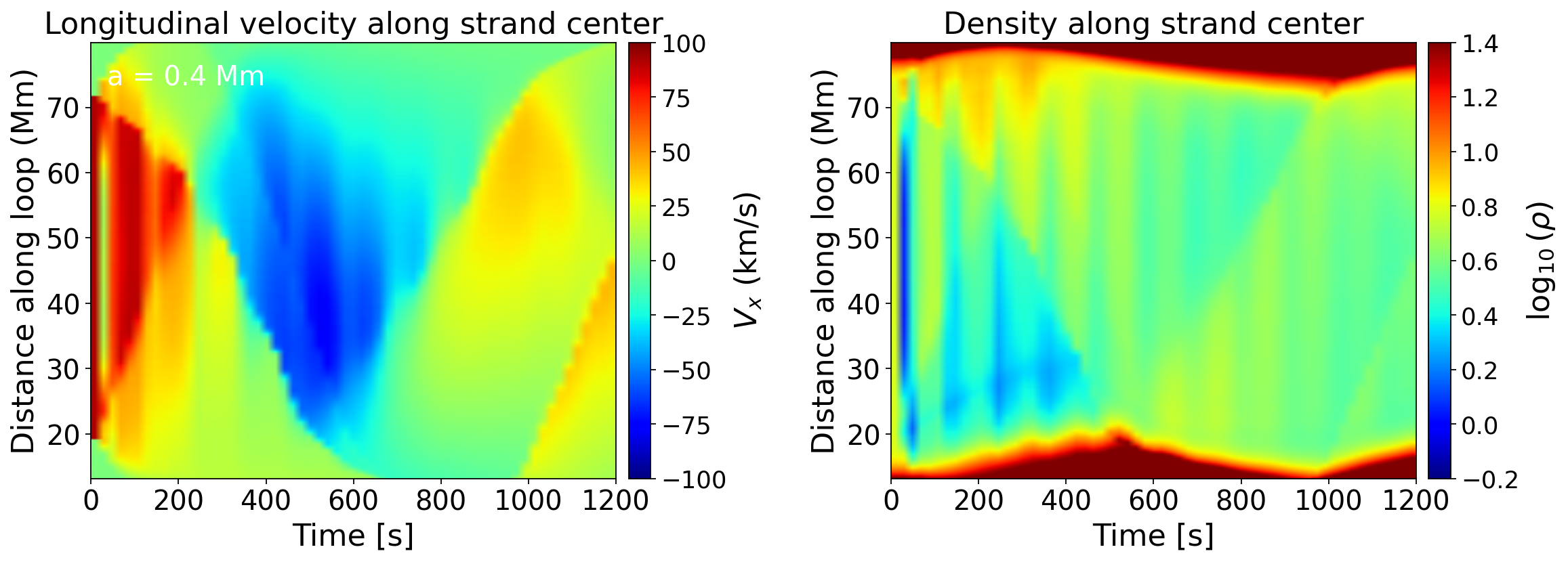}
 \includegraphics[width=13.0cm, height=4.0cm]{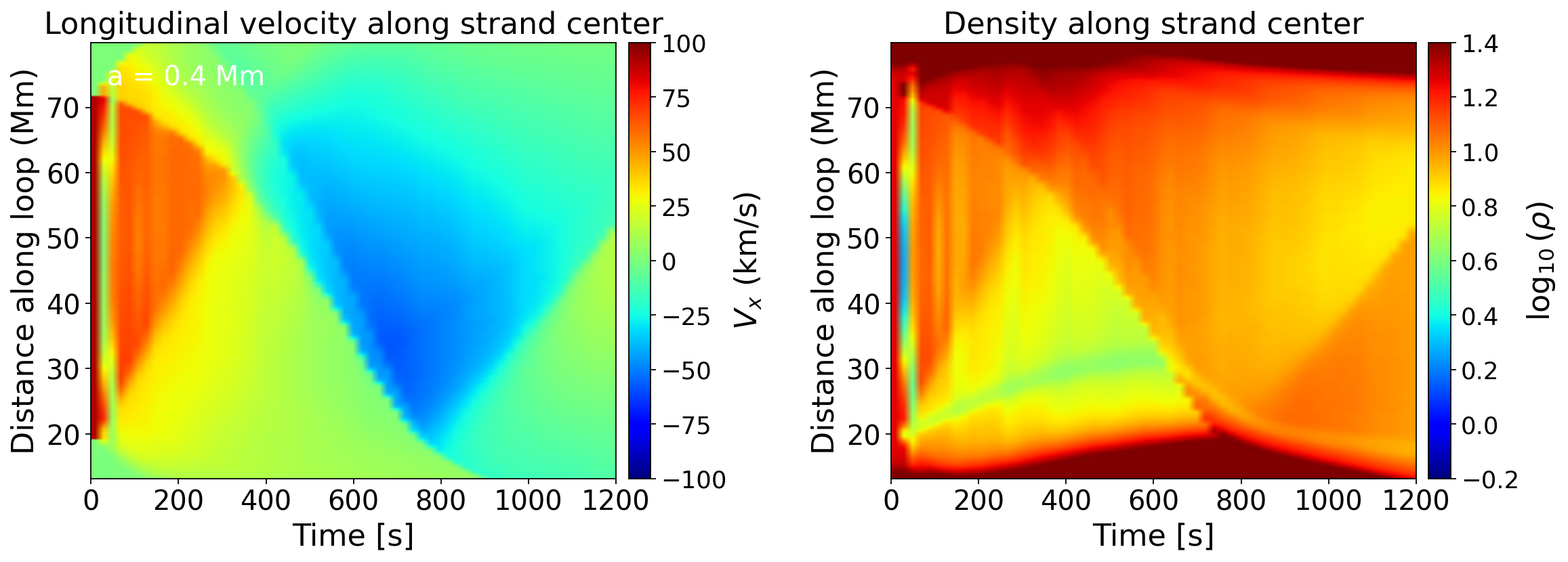}
\caption{Time--distance diagrams of longitudinal velocity $V_x$ (left column) and mass density (right column) along the strand axis(at z=0) for the bounded uniform flow case. The top and bottom rows correspond to density contrasts $d=5$ and $d=20$, respectively. The $V_x$ maps reveal the evolution of large-scale field-aligned flows that propagate, reflect, and periodically reverse direction along the loop. These motions are closely correlated with compressive disturbances evident in the density maps. The density panels clearly display nodal structures characteristic of standing slow-mode oscillations. The presence of multiple nodes and the shortened apparent wavelength relative to the loop length indicate higher-order harmonics, corresponding to approximately the second harmonic for $d=5$ and the third harmonic for $d=20$, rather than the fundamental standing mode.
}
\label{fig:time_distance_map}
\end{figure*}


\section{Results} 
\label{sec:Result}
In this work, we focus on impulsively excited vertically polarized kink oscillations in the tube-like coronal strands lying horizontally. To initiate these waves, we impose an initial perturbation on the vertical velocity component (V$_{z}$) \citep{2011ApJ...726...42S,2015A&A...582A..75O}. We deal with an isobaric strand with a density contrast, d=5 and d=20 times larger than the ambient coronal loop plasma considering uniform and non-uniform initial flow configurations (bounded, unbounded, and external) and the nonlinear longitudinal disturbances. We note that such a relatively low density contrast, d=5, is not strongly supported by Hinode/SOT observations; rather, observational studies \citep{2008A&A...482L...9O,2011ApJ...736..121A} indicate that significantly higher density contrasts (up to d=20) are more typical for fine strand-like structures. In this work, we include the case with d=5 primarily for comparison purposes with the higher contrast case (d=20), which is better aligned with the observational constraints. This comparison helps illustrate how varying the density contrast influences the physical properties and dynamics of the strand.

\subsection{Effect of the Initial Uniform Flow and Its Nonlinear Evolution on Kink Oscillations}

Figure~\ref{fig:fig5} illustrates the effect of different uniform flow configurations on the fundamental standing kink mode. The left panel shows a uniform  initial flow of $92~\mathrm{km~s^{-1}}$ localized within the strand width, consistent with Hinode/SOT observations\citep{2008A&A...482L...9O}. The middle panel depicts a flow of the same magnitude present both inside and outside the strand, while the right panel shows a flow existing only outside the strand.

Fig.~\ref{fig:fig2} presents the snapshots of the mass density ($\rho$), longitudinal velocity (V$_{x}$), and transverse velocity (V$_{z}$) at different stages of the simulation for the bounded, uniform case at density contrast, d=5. 
The evolution of the mass density, longitudinal velocity, and transverse velocity reveals the coupled dynamics of compressibility, field-aligned flows, and transverse wave motions in the system. At $t = 50$ s (first column), the strand is already undergoing kink oscillations triggered by the impulsive perturbation applied at $t = 0$. The longitudinal velocity $V_x$ shows localized compressive signatures at the strand boundaries, indicating coupling between transverse displacement and compressibility. The transverse velocity $V_z$ shows signatures extending beyond the strand into the surrounding plasma, indicating the onset of a global wave motion. The mass density $\rho$ primarily reflects compressive perturbations induced by the transverse kink oscillation during the coherent phase ($t \approx 200$ s), it exhibits mild enhancement and structuring within the strand, arising from periodic compression and rarefaction as the strand undergoes transverse displacement. This enhancement remains modest, consistent with the dominance of fast magnetoacoustic kink motion with weak compressibility. As the oscillation damps ($t \approx 400$ s), the density perturbations become smoother and less pronounced, indicating reduced compressive effects, and by the relaxation stage ($t \approx 1000$ s), only weak residual structuring persists.

The longitudinal velocity $V_x$ reflects these compressive dynamics along the magnetic field: it is strongest and localized near the strand boundaries during the coherent phase, where transverse motions induce pressure gradients and drive field-aligned flows. It weakens and becomes more diffuse as the kink oscillation damps, indicating diminished coupling between transverse and longitudinal motions, and at late times, it survives only as weak residual flows associated with the gradual relaxation of the system. In contrast, the transverse velocity $V_z$ captures both the localized kink oscillation and the broader global wave response during the coherent phase. It exhibits a clear dipolar structure across the strand, characteristic of kink motion, while simultaneously extending across the surrounding plasma as a spatially pattern, indicating the presence of a global wave motion generated by the impulsive perturbation. As the system evolves ($t \approx 400$ s), the amplitude of the localized kink motion decreases and the $V_z$ field becomes increasingly diffuse, reflecting the damping of both the strand oscillation and the global wave component. By $t \approx 1000$ s, the dipolar signature associated with kink motion largely disappears, leaving behind weak, spatially extended velocity patterns characteristic of a decaying large-scale wave field. Therefore, these diagnostics demonstrate that the impulsive excitation generates a coupled response consisting of a localized, compressive kink oscillation of the dense strand embedded within a broader global wave field, with energy progressively transferred and dissipated through compressive effects and field-aligned flows.

Figure~\ref{fig:fig3a} demonstrates the spatial evolution of the longitudinal plasma flow (V$_{x}$) in comparison with the local sound speed (c$_{s}$) for density contrasts d=5. The flow is introduced only as an initial condition and is not maintained by any external forcing or boundary driving. As time progresses, the initially imposed flow does not remain steady. Instead, its magnitude gradually decreases, redistributes along the loop, and even reverses direction at later times. This clearly indicates that the flow evolves self-consistently according to the MHD equations under open boundary conditions. The comparison with the local sound speed shows that the flow can initially be subsonic or near-sonic depending on the density contrast, but it weakens over time and becomes dynamically less dominant as the system evolves. This evolution demonstrates that the plasma flow is not artificially sustained in the model. Rather, it interacts with wave motions and undergoes natural redistribution, which can influence the damping and propagation of kink oscillations. A similar qualitative behavior is observed for both density contrast cases (d=5 and d=20), although the evolution is more pronounced in the higher density contrast case due to the lower internal sound speed.

We emphasize that our model is two-dimensional in the x–z plane with $\frac{\delta}{\delta y}$ = 0, following Gruszecki et al. 2008 \citep{2008A&A...487L..17V}. In this geometry, the Alfv\'en wave is decoupled and absent from the system, which supports only fast and slow magnetoacoustic modes (V${_y}$ = B${_y}$ = 0 throughout). The transverse oscillation reported here is therefore a fast magnetoacoustic kink mode of the dense strand, not an Alfv\'en wave. We identify and quantify the kink mode through the coherent transverse displacement of the dense strand itself (Fig.~\ref{fig:g1_vz}, top panel), rather than through the ambient velocity field, following Selwa et al. 2005\citep{2005A&A...440..385S}. At each instant we fit the transverse density profile across the strand with a Gaussian (cf., Fig.~\ref{fig:g1_vz}, top panel),
\begin{equation}
G(z) = G_0 \exp\!\left[-\frac{(z - G_1)^2}{2G_2^2}\right] + G_3,
\label{eq:gaussian_fit}
\end{equation}
where $G_1$ and $G_2$ denote the position of the strand centre and its half-width, respectively, while $(G_0 + G_3)$ gives the density at the strand center. The fitted center $G_1(t)$ traces the transverse motion of the strand axis and executes a damped oscillation at the fundamental kink period ($P \approx 108$~s for $d=5$). This demonstrates that the dense strand, not the entire simulation domain, oscillates as a coherent kink mode.
 
To quantify the damping oscillation, we fit the displacement-time profile of the strand (Fig.~\ref{fig:g1_vz}) using a damped sine function:

\begin{equation}
A(t) = A_{0} + A_{\rm max}\sin\left( \frac{2\pi t}{p} + \phi \right)\exp(-t/\tau),
\label{fitting_eqn}
\end{equation}

where, $A_{0}$ is the equilibrium position, $A_{\rm max}$ the amplitude, $p$ the period, $\tau$ the damping time, and $\phi$ the initial phase \citep{2005A&A...440..385S}. The fitted parameters for different flow conditions are summarized in Table~\ref{tab:uniformflow}.

We note that the transverse velocity maps display signatures that appear spatially extended across the computational domain (Fig.~\ref{fig:g1_vz}, bottom panel). However, this does not imply that the oscillations are purely global. Similar behavior has been reported by Selwa et al. (2005) \citep{2005A&A...440..385S}, where impulsively excited disturbances generate a wave packet consisting of both global loop oscillations and localized kink oscillations. In such systems, the initial perturbation excites a mixture of fast magnetoacoustic modes, including both the global oscillatory response of the surrounding plasma and the coherent transverse displacement of the dense strand. The latter corresponds to the kink mode, which is most robustly identified through the motion of the strand axis itself. In our simulations, the kink nature of the oscillation is confirmed by tracking the transverse displacement of the strand using Gaussian fitting of the density profile (Eq.~\ref{eq:gaussian_fit}). The resulting displacement exhibits a coherent, damped oscillation at the fundamental kink period, demonstrating that the dense strand oscillates as a distinct structure embedded within a larger-scale wave field. Therefore, the apparently global transverse velocity patterns seen in the $V_z$ maps (Fig.~\ref{fig:g1_vz}, bottom panel) are interpreted as part of the broader wave response to the initial perturbation, while the kink mode is more appropriately diagnosed through the localized displacement of the strand. This interpretation is consistent with previous numerical studies of impulsively excited loop oscillations, where multiple wave components coexist and the kink mode is best identified through the displacement of the waveguide itself \citep{2005A&A...440..385S}.

\subsubsection{Slow-Mode Response and Shock Formation}
Fig~\ref{fig:density_5} Left-panel, first-row, illustrates the time signature of the mass density evaluated at the middle of the strand corresponding to no flow i.e., $V = 0$. The second and third of the left-panels correspond to the cases when $V = 92$ km s$^{-1}$ is localized either within the width of the strand or existing both inside and outside the strand respectively. The fourth row of left panel of it corresponds to the specific situation when flow of the same magnitude is pervaded throughout the surrounding region of a strand but does not exist inside the width of the strand.

Kink oscillations are clearly seen in all left-panels. Mass density at the middle of the strand always decreases with time, as can be seen in the top left and bottom right panels of Fig.~\ref{fig:density_5}. This evacuation of the strand over time is due to the leakage of energy. During this process plasma flows into the surrounding region, smoothing the sharp edges of the strand and broadening in the width of the strand, as can be seen in Fig.~\ref{fig:density_5}. In our model, the ambient medium refers to the plasma surrounding the strand within the hosting coronal loop, rather than the large-scale diffuse corona. As noted in previous studies \citep{2008A&A...489..413G}, the properties of surrounding plasma layers can significantly influence wave behavior and structuring. In the present simulations, the interaction between the strand and its immediate environment appears to facilitate cross-boundary plasma diffusion, which causes the strand to remain broader even after wave amplitudes have decayed. We acknowledge that this behavior differs from the results of \citep{2005A&A...440..385S}, where the strand recovered its initial width after oscillation damping. This difference is likely related to variations in model setup, equilibrium profiles, numerical diffusion, or boundary treatment between the two studies. The red curves in right-panel of the Fig.~\ref{fig:density_5} illustrate the variation in different situations of the flow. A smooth variation in the half-width of the strand is discernible at the time of compression and rarefaction of the strand, which is out of phase with variation in mass density (black-curves in the same panel). Period of the slow mode wave is around 400 s. As the sound speed inside the strand is given by \(c_{Si} = \frac{c_{Se}}{\sqrt{d}} = 112~km s^{-1}\). Therefore, the period of the fundamental mode of the standing waves as estimated  \(\frac{2L_{sd}}{c_{Si}} = 1270{\rm s}\), where \(L_{sd}\) = 71 Mm is the length of the strand. 

Fig.~\ref{fig:density_contrast20}, left-column shows a comparison in the variation of mass density at the middle of the strand. Further, in order to explain the existence of slow sausage mode waves, Fig.~\ref{fig:density_contrast20}, right-column, displays the variation of the half-width of the strand at the middle of the strand as a function of time. Analogous to the previous findings, in this case we also get similar results, except in this case we treat the slow sausage wave as a slow shock.There are additional compression and rarefaction in the mass-density, as can be noticed in the second and third columns of the left panel in Fig.~\ref{fig:density_contrast20}. These are related to slow mode waves which are excited due to reflection of the plasma from the dense chromospheric layer. The existence of the slow mode waves can be easily illustrated by displaying the half-width of the strand at the center of the strand with respect to time.The reason for its interpretation as a slow shock can be understood in terms of the considered flow speed of 92 km s$^{-1}$, which is 1.64 time higher than the sound speed $c_{Si} = \frac{c_{Se}}{\sqrt{d}} = 56$ km s$^{-1}$ inside the strand, and therefore, can be considered as a supersonic flow. We believe that this slow mode wave is basically excited due to the flow itself, as we do not see a significant compression/rarefaction in mass density when the flow is absent, even though the pulse is present in the system (cf., top panel of top-left panel of Fig.~\ref{fig:density_contrast20}).

To assess the influence of plasma flows on kink oscillation properties, we estimate the variation in amplitude, period, and damping time relative to the no-flow case. The parameters of kink wave characteristics in different conditions of uniform flow are given in Table~\ref{tab:uniformflow}. For the density contrast d=5, the amplitude shows negligible variation (<1\%) across all flow conditions, indicating that flows do not significantly affect the oscillation amplitude across all cases and agree with Hinode/SOT observations \citep{2011ApJ...736..121A}. However, the period exhibits a slight increase of about~1\% in the presence of flows. In contrast, the damping time shows a more pronounced reduction, decreasing by approximately 20–25\% in the case of unbounded and external flows compared to the no-flow condition.
For higher density contrast, d=20, the amplitude remains nearly unchanged (<1\%) as reported \citep{2011ApJ...736..121A}, while the period varies mildly (~1–2\%). Notably, the damping time decreases significantly (~10–15\%) in the presence of flows, particularly for unbounded flow conditions consistent with Hinode/SOT measurements \citep{2008A&A...482L...9O, 2011ApJ...736..121A}. Although absolute variations appear small, percentage analysis reveals that damping time is the most sensitive parameter to plasma flows. Overall, these results suggest that while plasma flows have a minimal effect on the oscillation amplitude and period, they play a crucial role in enhancing the damping of kink oscillations. Furthermore, the reduction in damping time is more significant in cases with stronger or more extended flow profiles, indicating that flow-induced effects contribute to more efficient energy dissipation.

A comparison between these two cases, i.e., density contrast $d$ being equal to 5 and 20, shows that \(P\), \(\tau\), and \(A_{\max}\) of the kink waves are found to be higher in strand with high density contrast, i.e., $d=20$ as compared to the low-density contrast strand ($d=5$). However, the opposite trend is noticed in the value of \(\frac{\tau}{P}\). The increase in the value of the period corresponding to high density contrast isobaric strand is due to the decrease of the Alfv\'en speed inside the strand, whereas the increment in the value of \(\ A_{\max}\) is due to deposition of more energy in dense strand as compared to less dense strand by a given pulse of fixed strength and width. These results are analogous to the results of Selwa et al. (2006)\citep{2006A&A...454..653S} where they had also noticed the same behavior but without the effect of the flow. Further, the stronger damping (lower value of \(\frac{\tau}{P}\)) of the kink wave in the low-density contrast strand compared to high density contrast strand can be explained in terms of the large leakage of waves as reported by Selwa et al. (2007)\citep{2007A&A...462.1127S}. Overall, we interpret the slow sausage perturbation as a weak slow-mode shock. This is justified by the supersonic nature of the internal flow (Mach number $\sim$ 1.64), which facilitates nonlinear steepening of compressive slow-mode waves, leading to shock formation. The enhanced and self-consistent density compression signatures observed in the simulations further support this interpretation.

Moreover, in order to further confirm our argument regarding the generation of slow mode waves (Fig.~\ref{fig:unbounded_nopulse}), we re-analyze the time signature of the mass density and half-width of strand at the middle of the strand corresponding to un-bounded flow when pulse is not present in the system. The period of the slow sausage mode is $\sim$ 600s, which is 4 times shorter than the period of the fundamental mode, $\frac{2L_{sd}}{c_{Si}} = 2535$s, and is therefore more likely related to the third harmonic of the standing slow-mode wave\citep{2019A&A...632A..64D}. To further clarify the nature of the compressive perturbations and identify the dominant harmonic, we construct time-distance maps of the longitudinal velocity ($V_x$) (left-column) and mass density (right column) along the strand (Fig.~\ref{fig:time_distance_map}) in case of bounded uniform flow. These diagnostics allow us to distinguish between propagating, standing, and nonlinear wave behavior. The maps display clear alternating ridges of enhanced and reduced amplitude, persisting over multiple reflections between the loop footpoints, which confirms the formation of \textit{standing slow-mode waves}, rather than propagating or sloshing motions. The spatial distribution reveals multiple nodes and anti-nodes along the loop, indicating the presence of higher harmonics. The inferred wavelength is significantly shorter than twice the loop length, ruling out the fundamental mode. For the d=5 case (Fig.~\ref{fig:time_distance_map}, top row), the node structure is consistent with the second harmonic, whereas for d = 20, a shorter spatial scale and more complex pattern indicate dominance of the third harmonic (Fig.~\ref{fig:time_distance_map}, bottom row). This is consistent with the stronger compressive response at higher density contrast. Furthermore, the phase relation between density and longitudinal velocity confirms the compressive nature of slow magnetoacoustic waves. The absence of strong discontinuities suggests predominantly weakly nonlinear behavior, although mild steepening is visible for d=20. Therefore, Fig.~\ref{fig:time_distance_map} demonstrates that the observed periodicities correspond to higher harmonic standing slow-mode waves.

\begin{figure*}[h]
\centering
\includegraphics[width=\textwidth]{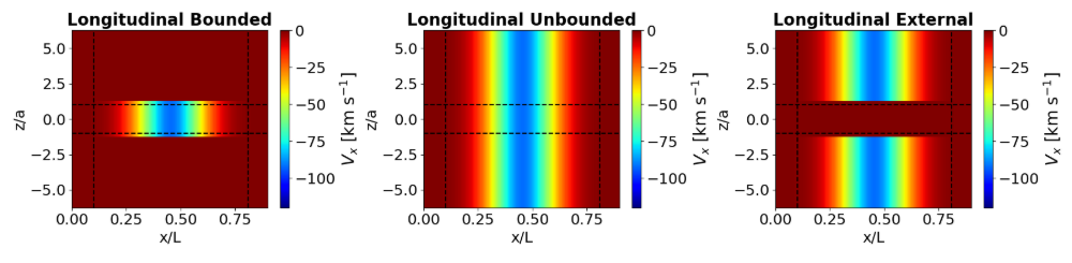}
\caption{Initial spatial distribution of the longitudinal velocity $V_{x}$ (km~s$^{-1}$) for the three longitudinally varying (Gaussian) flow configurations, shown in the $x$--$z$ plane. The flow follows the Gaussian profile of Eq.~\ref{eq:longitudinal_flow} with peak amplitude $v_{0} = -92$~km~s$^{-1}$ centred at $x_{0} = 0.455\,L$. Left: \emph{Longitudinal Bounded} flow, with the Gaussian profile confined to the strand interior. Middle: \emph{Longitudinal Unbounded} flow, where the Gaussian profile extends across the full $z$-domain, both inside the strand and throughout the ambient corona. Right: \emph{Longitudinal External} flow, where the Gaussian profile is imposed in the ambient corona but excluded from the strand interior. The horizontal dashed lines indicate the strand boundaries at $z/L = \pm 0.002$, i.e.\ $z = \pm 0.2$~Mm, while the vertical dashed lines mark the chromospheric layers at $x = 0.10\,L$ and $x = 0.81\,L$.}
\label{fig:nonuniformflows}
\end{figure*}

\begin{figure*}[h]
\includegraphics[width=0.33\textwidth, height=0.15\textheight, keepaspectratio=false]{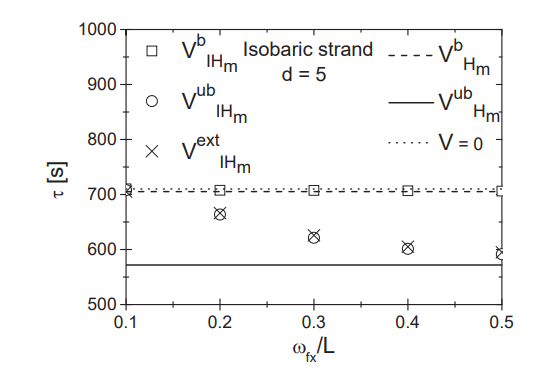}
\includegraphics[width=0.33\textwidth, height=0.15\textheight, keepaspectratio=false]{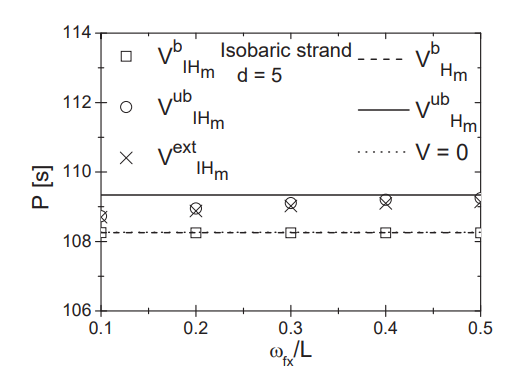}
\includegraphics[width=0.33\textwidth, height=0.15\textheight,keepaspectratio=false]{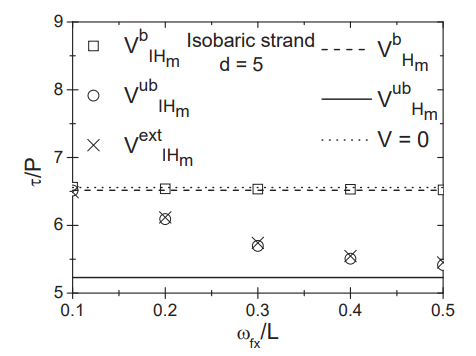}

\includegraphics[width=0.33\textwidth, height=0.15\textheight, keepaspectratio=false]{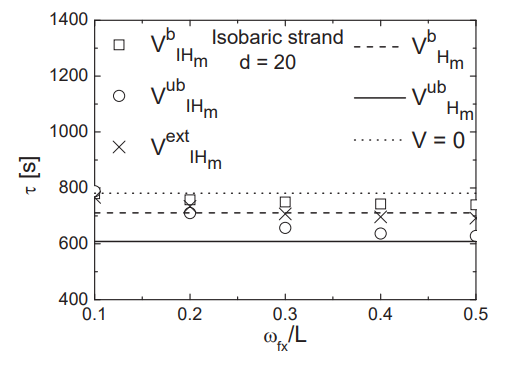}
\includegraphics[width=0.33\textwidth, height=0.15\textheight, keepaspectratio=false]{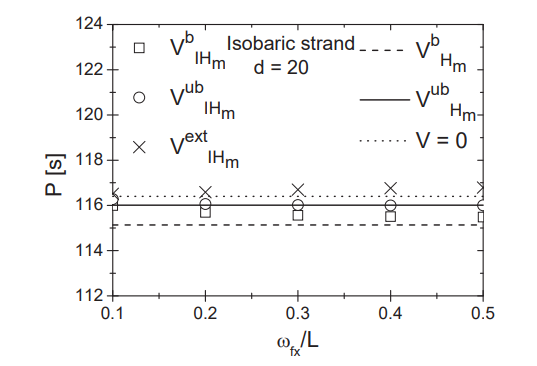}
\includegraphics[width=0.33\textwidth, height=0.15\textheight, keepaspectratio=false]{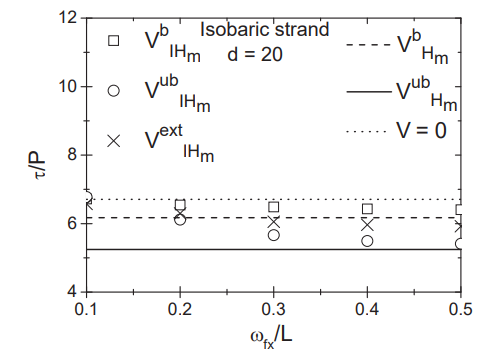}

\caption{Variation of the damping time \(\tau\) (left column), period \(P\) (middle column), and the ratio \(\tau/P\) (right column) of the fundamental mode of standing kink waves as a function of the normalized half-width of a Gaussian flow with a uniform profile. The top and bottom row respectively, corresponds to an isobaric strand with a density contrast of 5 and 20 relative to the ambient coronal loop plasma.}
\label{fig:kinkwave_variation}
\end{figure*}

\subsubsection{Damping Behavior of Kink Waves in the Presence of Longitudinally Varying Inhomogeneous Flows}

To investigate how longitudinally varying initial flow profiles and the nonlinear disturbances influence the damping of standing kink oscillations, we consider three distinct Gaussian initial-flow configurations, illustrated in Fig.~\ref{fig:nonuniformflows}. In the first case, the flow is confined strictly within the strand width (left panel), hereafter denoted as $V_{\mathrm{IH}_m}^{b}$. The second configuration extends the same Gaussian flow both inside and outside the strand, producing an unbounded profile ($V_{\mathrm{IH}_m}^{ub}$; middle panel). The third configuration removes the flow from the strand interior entirely while allowing it to persist throughout the surrounding corona ($V_{\mathrm{IH}_m}^{ext}$; right panel). The influence of these inhomogeneous flow patterns on the kink-mode properties is quantified by measuring the damping time $\tau$, period $P$, and quality factor $\tau/P$ as functions of the normalized flow half-width $\omega_{fx}/L$. The corresponding results for density contrasts $d = 5$ and $d = 20$ are presented in the top and bottom panels of Fig.~\ref{fig:kinkwave_variation}, respectively.

We find that both the damping time $\tau$ and quality factor $\tau/P$ decrease systematically with increasing flow width. Among the three flow geometries, the unbounded configuration ($V_{\mathrm{IH}_m}^{ub}$) produces the strongest reduction in damping time, indicating enhanced coupling between the oscillating strand and the external plasma. In contrast, both the bounded ($V_{\mathrm{IH}_m}^{b}$) and external-only ($V_{\mathrm{IH}_m}^{ext}$) configurations yield significantly weaker modifications to the damping rate. It is important to note that initially uniform flows become inhomogeneous as they partially reflect from the dense chromospheric layer. As a result, inhomogeneous flows with sufficiently large half-widths naturally converge toward the damping behavior of the uniform-flow limit. This convergence occurs more rapidly for strands with higher density contrast. For example, increasing $\omega_{fx}/L$ from 0.2 to 0.4 in the unbounded case reduces the damping time by approximately 9.4\% for $d=5$, but by 10.3\% for $d=20$. In the bounded-flow configuration, the reduction in $\tau$ is negligible for $d=5$ and only $\sim 1.9\%$ for $d=20$, reinforcing that the external flow component plays a dominant role in enhancing damping. The oscillation period $P$ remains largely insensitive to the flow width in all cases, consistent with the expectation that longitudinal flow structuring predominantly affects dissipative properties rather than the fundamental kink-mode frequency.

\begin{figure*}[h]
\centering
\includegraphics[width=0.4\textwidth, height=0.2\textheight, keepaspectratio=false]{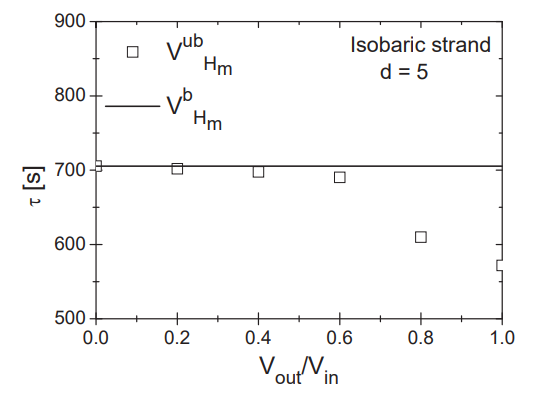}
\includegraphics[width=0.4\textwidth, height=0.2\textheight, keepaspectratio=false]{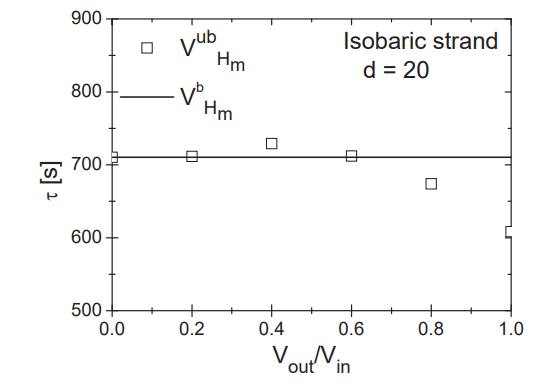}

\caption{Variation of the damping time ($\tau$) of the fundamental mode of the standing kink wave as a function of the ratio of the amplitudes of two uniform flows inside and outside the strand. The left panel corresponds to an isobaric strand with a density contrast of 5, the right panel to a contrast of 20.}
\label{fig:init_profiles}
\end{figure*}

\subsection{Different Flows Inside and Outside the strand}

We explore the influence of uniform flow magnitudes inside and outside the strand on the damping of standing kink oscillations. Unlike previous cases where the flow amplitude was uniform throughout the domain, here we introduce two distinct uniform initial flows: one confined within the strand and another occupying the external region. This configuration mimics realistic coronal loop conditions where plasma flows can differ significantly between the dense strand core and its surrounding corona. The internal flow amplitude is denoted by $V_{\rm in}$, while the external flow amplitude is $V_{\rm out}$. We systematically vary the ratio $V_{\rm out}/V_{\rm in}$ from 1.0 (equal flows) to values approaching zero (dominant internal flow) to assess its impact on wave damping. Figure~\ref{fig:init_profiles} illustrates the variation of damping time $\tau$ as a function of this ratio for an isobaric strand $d=5$ (left panel) and $d=20$ (right panel) .

Our results reveal a clear trend: when $V_{\rm out}/V_{\rm in} = 1.0$, the damping time is minimal, it indicates the strongest damping of kink oscillations. This behavior can be attributed to enhanced energy leakage facilitated by symmetric shear at the strand boundaries, which promotes efficient coupling between the oscillating strand and the surrounding plasma. As the ratio decreases, the damping time $\tau$ increases approximately linearly until $V_{\rm out}/V_{\rm in} \approx 0.5$. Beyond this threshold, further reduction in external flow amplitude produces negligible changes in $\tau$, and the damping time approaches the value corresponding to a bounded internal flow. Physically, this saturation effect suggests that once the external flow becomes sufficiently weak, its contribution to shear-driven energy dissipation diminishes, leaving the internal flow as the dominant factor. The oscillation period $P$ remains largely unaffected across all ratios, confirming that flow asymmetry primarily influences damping rather than wave frequency. The quality factor $\tau/P$ follows the same trend as $\tau$, reinforcing the conclusion that damping efficiency depends strongly on the relative strength of internal and external flows. 
 
\section{Discussion and Conclusions} \label{sec:D&C}

The present study investigates the influence of the nonlinear longitudinal disturbances triggered by field-aligned initial flows on the damping of fundamental standing kink waves in solar coronal loops, focusing exclusively on isobaric strands.  Using two-dimensional numerical simulations, we analyzed how the nonlinear disturbances generated by different initial flow profiles—bounded, unbounded, and external—affect wave properties such as period, damping time, and quality factor. Our results provide new insights into the interaction between plasma flows and kink oscillations in coronal loops. These findings have broader implications for coronal heating and energy transport mechanisms. We found enhanced damping caused by unbounded flows suggests efficient energy dissipation, which may contribute to localized heating in coronal loops. Furthermore, accurate numerical simulation of flow-induced damping improves coronal seismology diagnostics, enabling better estimates of magnetic field strength and coronal seismology.

Previously, Gruszecki et al. (2008) \citep{2008A&A...488..757G} examined how bounded field-aligned flows (confined within a 2.5~Mm strand) modify standing fast magnetoacoustic kink modes, showing that such flows can shift the oscillation period and alter mode structure. In contrast, our study employs realistic isobaric strands with strong observational density contrasts (d = 5, 20; width 0.4 Mm) and systematically explores multiple initial-flow geometries—bounded, unbounded, external, and longitudinally inhomogeneous; crucially, whereas those earlier models assumed steady flows, the field-aligned flow imposed here is an initial condition that does not remain steady but evolves nonlinearly into longitudinal disturbances. We find that the geometry of the initial flow, rather than amplitude alone, is the dominant factor controlling damping: the nonlinear disturbances triggered by unbounded and inhomogeneous initial flows reduce the damping time by up to $\sim 25\%$ and excite additional slow-mode and slow-shock components. When flows extend beyond the strand width, the damping becomes significantly stronger, indicating that large-scale flow structures interact more efficiently with kink waves. These results highlight the crucial role of extended flow profiles in energy dissipation and forward modelling of cool, fine-structured coronal loops.

We also find that the presence of flow introduces additional damping compared to the no-flow case. The reduction in damping time is linked to the scattering of kink waves by the nonlinear longitudinal disturbances and their inhomogeneities.  When the scale of the flow variation approaches the wavelength of the fundamental mode, the damping becomes much stronger. This highlights the importance of considering flow geometry and localization in models of coronal loop oscillations. Furthermore, the excitation of slow-mode components in some cases suggests that flows can generate mixed-mode wave packets, consistent with previous theoretical predictions. To validate our simulation, we compare the damping times obtained for an isobaric strand with observational ranges reported by Hinode/SOT and SDO/AIA, reported by \citep{2019ApJS..241...31N, 2011ApJ...736..121A}. We found that our results lie within the observed range of $500$--$1500$ s, confirming the consistency of our model with realistic coronal loop oscillations.

In conclusion, our results emphasize that flow patterns particularly unbounded and inhomogeneous flows, are key factors in the damping of kink oscillations in isobaric strands. These findings have important implications for coronal seismology and the interpretation of observational data. Future work should extend this analysis to non-isobaric conditions with non-uniform flows. High-resolution observations from missions such as Solar Orbiter will be essential to validate these predictions and improve our understanding of wave dynamics and energy transport mechanisms in the solar corona. Future work should incorporate additional physical effects such as gravity, radiative losses, and thermal conduction to capture non-isothermal conditions. Extending the model to three dimensions and including realistic magnetic field curvature will further improve its applicability to observed coronal loop dynamics.

\ack{We sincerely thank both referees for their constructive and valuable comments that significantly improved the manuscript. Balveer Singh acknowledges the project titled “100 years of the global solar corona: new insights from historical data”, which is funded by the Science and Technology Facilities Council (STFC), United Kingdom Research and Innovation (UKRI) through grant ST/W001098/1. Balveer Singh sincerely acknowledges Karen Meyer for valuable discussions that greatly contributed to this project. This work was supported by the Science and Engineering Research Board-Department of Science and Technology, Core Research Grant (SERB-DST, CRG) Project, Grant No. (CRG/2022/007017). Shakti Singh acknowledges financial support from the Council of Scientific and Industrial Research (CSIR), India, under File No. 09/1143(15987)/2022-EMR-I. The authors acknowledge the use of the ATHENA code and of IDL \& PYTHON libraries for numerical data analysis.}
\begin{fmtext}
\end{fmtext}

\bibliographystyle{vancouver}
\bibliography{sample}

@ARTICLE{2010ApJ...716..154A,
       author = {{Antolin}, P. and {Shibata}, K. and {Vissers}, G.},
        title = "{Coronal Rain as a Marker for Coronal Heating Mechanisms}",
      journal = {\apj},
         year = 2010,
        month = jun,
       volume = {716},
       number = {1},
        pages = {154-166},
          doi = {10.1088/0004-637X/716/1/154},
archivePrefix = {arXiv},
       eprint = {0910.2383},
 primaryClass = {astro-ph.SR},
       adsurl = {https://ui.adsabs.harvard.edu/abs/2010ApJ...716..154A}
}

@ARTICLE{2011ApJ...736..121A,
       author = {{Antolin}, P. and {Verwichte}, E.},
        title = "{Transverse Oscillations of Loops with Coronal Rain Observed by Hinode/Solar Optical Telescope}",
      journal = {\apj},
         year = 2011,
        month = aug,
       volume = {736},
       number = {2},
          eid = {121},
        pages = {121},
          doi = {10.1088/0004-637X/736/2/121},
archivePrefix = {arXiv},
       eprint = {1105.2175},
 primaryClass = {astro-ph.SR},
       adsurl = {https://ui.adsabs.harvard.edu/abs/2011ApJ...736..121A}
}

@ARTICLE{2005ApJ...624L..57A,
       author = {{Andries}, Jesse and {Arregui}, Inigo and {Goossens}, Marcel},
        title = "{Determination of the Coronal Density Stratification from the Observation of Harmonic Coronal Loop Oscillations}",
      journal = {\apjl},
         year = 2005,
        month = may,
       volume = {624},
       number = {1},
        pages = {L57-L60},
          doi = {10.1086/430347},
       adsurl = {https://ui.adsabs.harvard.edu/abs/2005ApJ...624L..57A}
}

@ARTICLE{2003ApJ...598.1375A,
       author = {{Aschwanden}, Markus J. and {Nightingale}, Richard W. and {Andries}, Jesse and {Goossens}, Marcel and {Van Doorsselaere}, Tom},
        title = "{Observational Tests of Damping by Resonant Absorption in Coronal Loop Oscillations}",
      journal = {\apj},
         year = 2003,
        month = dec,
       volume = {598},
       number = {2},
        pages = {1375-1386},
          doi = {10.1086/379104},
archivePrefix = {arXiv},
       eprint = {astro-ph/0309470},
 primaryClass = {astro-ph},
       adsurl = {https://ui.adsabs.harvard.edu/abs/2003ApJ...598.1375A}
}

@ARTICLE{2011ApJ...732...81A,
       author = {{Aschwanden}, Markus J. and {Boerner}, Paul},
        title = "{Solar Corona Loop Studies with the Atmospheric Imaging Assembly. I. Cross-sectional Temperature Structure}",
      journal = {\apj},
         year = 2011,
        month = may,
       volume = {732},
       number = {2},
          eid = {81},
        pages = {81},
          doi = {10.1088/0004-637X/732/2/81},
archivePrefix = {arXiv},
       eprint = {1103.0228},
 primaryClass = {astro-ph.SR},
       adsurl = {https://ui.adsabs.harvard.edu/abs/2011ApJ...732...81A}
}

@ARTICLE{1999ApJ...520..880A,
       author = {{Aschwanden}, Markus J. and {Fletcher}, Lyndsay and {Schrijver}, Carolus J. and {Alexander}, David},
        title = "{Coronal Loop Oscillations Observed with the Transition Region and Coronal Explorer}",
      journal = {\apj},
         year = 1999,
        month = aug,
       volume = {520},
       number = {2},
        pages = {880-894},
          doi = {10.1086/307502},
       adsurl = {https://ui.adsabs.harvard.edu/abs/1999ApJ...520..880A}
}

@ARTICLE{2008ApJ...689L..73C,
       author = {{Chae}, Jongchul and {Ahn}, Kwangsoo and {Lim}, Eun-Kyung and {Choe}, G.~S. and {Sakurai}, Takashi},
        title = "{Persistent Horizontal Flows and Magnetic Support of Vertical Threads in a Quiescent Prominence}",
      journal = {\apjl},
         year = 2008,
        month = dec,
       volume = {689},
       number = {1},
        pages = {L73},
          doi = {10.1086/595785},
       adsurl = {https://ui.adsabs.harvard.edu/abs/2008ApJ...689L..73C}
}

@ARTICLE{2019A&A...632A..64D,
       author = {{Duckenfield}, T.~J. and {Goddard}, C.~R. and {Pascoe}, D.~J. and {Nakariakov}, V.~M.},
        title = "{Observational signatures of the third harmonic in a decaying kink oscillation of a coronal loop}",
      journal = {\aap},
         year = 2019,
        month = dec,
       volume = {632},
          eid = {A64},
        pages = {A64},
          doi = {10.1051/0004-6361/201936822},
       adsurl = {https://ui.adsabs.harvard.edu/abs/2019A&A...632A..64D}
}

@ARTICLE{2005JCoPh.205..509G,
       author = {{Gardiner}, Thomas A. and {Stone}, James M.},
        title = "{An unsplit Godunov method for ideal MHD via constrained transport}",
      journal = {Journal of Computational Physics},
         year = 2005,
        month = may,
       volume = {205},
       number = {2},
        pages = {509-539},
          doi = {10.1016/j.jcp.2004.11.016},
archivePrefix = {arXiv},
       eprint = {astro-ph/0501557},
 primaryClass = {astro-ph},
       adsurl = {https://ui.adsabs.harvard.edu/abs/2005JCoPh.205..509G}
}

@ARTICLE{2019A&A...627A..62G,
       author = {{Gupta}, G.~R. and {Del Zanna}, G. and {Mason}, H.~E.},
        title = "{Exploring the damping of Alfv{\'e}n waves along a long off-limb coronal loop, up to 1.4 Solar Radii}",
      journal = {\aap},
         year = 2019,
        month = jul,
       volume = {627},
          eid = {A62},
        pages = {A62},
          doi = {10.1051/0004-6361/201935357},
archivePrefix = {arXiv},
       eprint = {1905.08194},
 primaryClass = {astro-ph.SR},
       adsurl = {https://ui.adsabs.harvard.edu/abs/2019A&A...627A..62G}
}

@ARTICLE{2008A&A...488..757G,
       author = {{Gruszecki}, M. and {Murawski}, K. and {Ofman}, L.},
        title = "{Standing fast magnetoacoustic kink waves of solar coronal loops with field-aligned flow}",
      journal = {\aap},
         year = 2008,
        month = sep,
       volume = {488},
       number = {2},
        pages = {757-761},
          doi = {10.1051/0004-6361:200809873},
       adsurl = {https://ui.adsabs.harvard.edu/abs/2008A&A...488..757G}
}

@ARTICLE{2008A&A...489..413G,
       author = {{Gruszecki}, M. and {Murawski}, K. and {McLaughlin}, J.~A.},
        title = "{Influence of a dense photosphere-like layer on vertical oscillations of a curved coronal slab}",
      journal = {\aap},
         year = 2008,
        month = oct,
       volume = {489},
       number = {1},
        pages = {413-418},
          doi = {10.1051/0004-6361:20078092},
       adsurl = {https://ui.adsabs.harvard.edu/abs/2008A&A...489..413G}
}

@ARTICLE{2002A&A...394L..39G,
       author = {{Goossens}, M. and {Andries}, J. and {Aschwanden}, M.~J.},
        title = "{Coronal loop oscillations. An interpretation in terms of resonant absorption of quasi-mode kink oscillations}",
      journal = {\aap},
         year = 2002,
        month = nov,
       volume = {394},
        pages = {L39-L42},
          doi = {10.1051/0004-6361:20021378},
       adsurl = {https://ui.adsabs.harvard.edu/abs/2002A&A...394L..39G}
}

@ARTICLE{2011Natur.475..477M,
       author = {{McIntosh}, Scott W. and {de Pontieu}, Bart and {Carlsson}, Mats and {Hansteen}, Viggo and {Boerner}, Paul and {Goossens}, Marcel},
        title = "{Alfv{\'e}nic waves with sufficient energy to power the quiet solar corona and fast solar wind}",
      journal = {\nat},
         year = 2011,
        month = jul,
       volume = {475},
       number = {7357},
        pages = {477-480},
          doi = {10.1038/nature10235},
       adsurl = {https://ui.adsabs.harvard.edu/abs/2011Natur.475..477M}
}

@ARTICLE{2003A&A...411..605M,
       author = {{M{\"u}ller}, D.~A.~N. and {Hansteen}, V.~H. and {Peter}, H.},
        title = "{Dynamics of solar coronal loops. I. Condensation in cool loops and its effect on transition region lines}",
      journal = {\aap},
         year = 2003,
        month = dec,
       volume = {411},
        pages = {605-613},
          doi = {10.1051/0004-6361:20031328},
       adsurl = {https://ui.adsabs.harvard.edu/abs/2003A&A...411..605M}
}

@ARTICLE{1995SoPh..159..213N,
       author = {{Nakariakov}, V.~M. and {Roberts}, B.},
        title = "{Magnetosonic Waves in Structured Atmospheres with Steady Flows, I}",
      journal = {\solphys},
         year = 1995,
        month = jul,
       volume = {159},
       number = {2},
        pages = {213-228},
          doi = {10.1007/BF00686530},
       adsurl = {https://ui.adsabs.harvard.edu/abs/1995SoPh..159..213N}
}

@ARTICLE{2001A&A...372L..53N,
       author = {{Nakariakov}, V.~M. and {Ofman}, L.},
        title = "{Determination of the coronal magnetic field by coronal loop oscillations}",
      journal = {\aap},
         year = 2001,
        month = jun,
       volume = {372},
        pages = {L53-L56},
          doi = {10.1051/0004-6361:20010607},
       adsurl = {https://ui.adsabs.harvard.edu/abs/2001A&A...372L..53N}
}

@ARTICLE{2021SSRv..217...73N,
       author = {{Nakariakov}, V.~M. and {Anfinogentov}, S.~A. and {Antolin}, P. and {Jain}, R. and {Kolotkov}, D.~Y. and {Kupriyanova}, E.~G. and {Li}, D. and {Magyar}, N. and {Nistic{\`o}}, G. and {Pascoe}, D.~J. and {Srivastava}, A.~K. and {Terradas}, J. and {Vasheghani Farahani}, S. and {Verth}, G. and {Yuan}, D. and {Zimovets}, I.~V.},
        title = "{Kink Oscillations of Coronal Loops}",
      journal = {\ssr},
         year = 2021,
        month = sep,
       volume = {217},
       number = {6},
          eid = {73},
        pages = {73},
          doi = {10.1007/s11214-021-00847-2},
archivePrefix = {arXiv},
       eprint = {2109.11220},
 primaryClass = {astro-ph.SR},
       adsurl = {https://ui.adsabs.harvard.edu/abs/2021SSRv..217...73N}
}

@ARTICLE{2008A&A...482L...9O,
       author = {{Ofman}, L. and {Wang}, T.~J.},
        title = "{Hinode observations of transverse waves with flows in coronal loops}",
      journal = {\aap},
         year = 2008,
        month = may,
       volume = {482},
       number = {2},
        pages = {L9-L12},
          doi = {10.1051/0004-6361:20079340},
       adsurl = {https://ui.adsabs.harvard.edu/abs/2008A&A...482L...9O}
}

@ARTICLE{2013A&A...556A.104P,
       author = {{Peter}, H. and {Bingert}, S. and {Klimchuk}, J.~A. and {de Forest}, C. and {Cirtain}, J.~W. and {Golub}, L. and {Winebarger}, A.~R. and {Kobayashi}, K. and {Korreck}, K.~E.},
        title = "{Structure of solar coronal loops: from miniature to large-scale}",
      journal = {\aap},
         year = 2013,
        month = aug,
       volume = {556},
          eid = {A104},
        pages = {A104},
          doi = {10.1051/0004-6361/201321826},
archivePrefix = {arXiv},
       eprint = {1306.4685},
 primaryClass = {astro-ph.SR},
       adsurl = {https://ui.adsabs.harvard.edu/abs/2013A&A...556A.104P}
}

@ARTICLE{2016A&A...589A.136P,
       author = {{Pascoe}, D.~J. and {Goddard}, C.~R. and {Nistic{\`o}}, G. and {Anfinogentov}, S. and {Nakariakov}, V.~M.},
        title = "{Coronal loop seismology using damping of standing kink oscillations by mode coupling}",
      journal = {\aap},
         year = 2016,
        month = may,
       volume = {589},
          eid = {A136},
        pages = {A136},
          doi = {10.1051/0004-6361/201628255},
       adsurl = {https://ui.adsabs.harvard.edu/abs/2016A&A...589A.136P}
}

@ARTICLE{2002ApJ...577..475R,
       author = {{Ruderman}, M.~S. and {Roberts}, B.},
        title = "{The Damping of Coronal Loop Oscillations}",
      journal = {\apj},
         year = 2002,
        month = sep,
       volume = {577},
       number = {1},
        pages = {475-486},
          doi = {10.1086/342130},
       adsurl = {https://ui.adsabs.harvard.edu/abs/2002ApJ...577..475R}
}

@ARTICLE{2007A&A...462.1127S,
       author = {{Selwa}, M. and {Murawski}, K. and {Solanki}, S.~K. and {Wang}, T.~J.},
        title = "{Energy leakage as an attenuation mechanism for vertical kink oscillations in solar coronal wave guides}",
      journal = {\aap},
         year = 2007,
        month = feb,
       volume = {462},
       number = {3},
        pages = {1127-1135},
          doi = {10.1051/0004-6361:20065122},
       adsurl = {https://ui.adsabs.harvard.edu/abs/2007A&A...462.1127S}
}

@ARTICLE{2006A&A...454..653S,
       author = {{Selwa}, M. and {Solanki}, S.~K. and {Murawski}, K. and {Wang}, T.~J. and {Shumlak}, U.},
        title = "{Numerical simulations of impulsively generated vertical oscillations in a solar coronal arcade loop}",
      journal = {\aap},
         year = 2006,
        month = aug,
       volume = {454},
       number = {2},
        pages = {653-661},
          doi = {10.1051/0004-6361:20054286},
       adsurl = {https://ui.adsabs.harvard.edu/abs/2006A&A...454..653S}
}

@ARTICLE{2005A&A...440..385S,
       author = {{Selwa}, M. and {Murawski}, K. and {Solanki}, S.~K. and {Wang}, T.~J. and {T{\'o}th}, G.},
        title = "{Numerical simulations of vertical oscillations of a solar coronal loop}",
      journal = {\aap},
         year = 2005,
        month = sep,
       volume = {440},
       number = {1},
        pages = {385-390},
          doi = {10.1051/0004-6361:20053121},
       adsurl = {https://ui.adsabs.harvard.edu/abs/2005A&A...440..385S}
}

@ARTICLE{2004A&A...427.1065T,
       author = {{Teriaca}, L. and {Banerjee}, D. and {Falchi}, A. and {Doyle}, J.~G. and {Madjarska}, M.~S.},
        title = "{Transition region small-scale dynamics as seen by SUMER on SOHO}",
      journal = {\aap},
         year = 2004,
        month = dec,
       volume = {427},
        pages = {1065-1074},
          doi = {10.1051/0004-6361:20040503},
       adsurl = {https://ui.adsabs.harvard.edu/abs/2004A&A...427.1065T}
}

@ARTICLE{2003SoPh..217..199T,
       author = {{Terra-Homem}, M. and {Erd{\'e}lyi}, R. and {Ballai}, I.},
        title = "{Linear and non-linear MHD wave propagation in steady-state magnetic cylinders}",
      journal = {\solphys},
         year = 2003,
        month = nov,
       volume = {217},
       number = {2},
        pages = {199-223},
          doi = {10.1023/B:SOLA.0000006901.22169.59},
       adsurl = {https://ui.adsabs.harvard.edu/abs/2003SoPh..217..199T}
}

@ARTICLE{2008ApJ...687L.115T,
       author = {{Terradas}, J. and {Andries}, J. and {Goossens}, M. and {Arregui}, I. and {Oliver}, R. and {Ballester}, J.~L.},
        title = "{Nonlinear Instability of Kink Oscillations due to Shear Motions}",
      journal = {\apjl},
         year = 2008,
        month = nov,
       volume = {687},
       number = {2},
        pages = {L115},
          doi = {10.1086/593203},
archivePrefix = {arXiv},
       eprint = {0809.3664},
 primaryClass = {astro-ph},
       adsurl = {https://ui.adsabs.harvard.edu/abs/2008ApJ...687L.115T}
}

@ARTICLE{2004A&A...421L..33W,
       author = {{Wang}, T.~J. and {Solanki}, S.~K.},
        title = "{Vertical oscillations of a coronal loop observed by TRACE}",
      journal = {\aap},
         year = 2004,
        month = jul,
       volume = {421},
        pages = {L33-L36},
          doi = {10.1051/0004-6361:20040186},
       adsurl = {https://ui.adsabs.harvard.edu/abs/2004A&A...421L..33W}
}

@ARTICLE{2001ApJ...553L..81W,
       author = {{Winebarger}, Amy R. and {DeLuca}, Edward E. and {Golub}, Leon},
        title = "{Apparent Flows above an Active Region Observed with the Transition Region and Coronal Explorer}",
      journal = {\apjl},
         year = 2001,
        month = may,
       volume = {553},
       number = {1},
        pages = {L81-L84},
          doi = {10.1086/320496},
       adsurl = {https://ui.adsabs.harvard.edu/abs/2001ApJ...553L..81W}
}

@ARTICLE{2002ApJ...567L..89W,
       author = {{Winebarger}, Amy R. and {Warren}, Harry and {van Ballegooijen}, Adriaan and {DeLuca}, Edward E. and {Golub}, Leon},
        title = "{Steady Flows Detected in Extreme-Ultraviolet Loops}",
      journal = {\apjl},
         year = 2002,
        month = mar,
       volume = {567},
       number = {1},
        pages = {L89-L92},
          doi = {10.1086/339796},
       adsurl = {https://ui.adsabs.harvard.edu/abs/2002ApJ...567L..89W}
}

@ARTICLE{2002ApJ...576L.153O,
       author = {{Ofman}, L. and {Aschwanden}, M.~J.},
        title = "{Damping Time Scaling of Coronal Loop Oscillations Deduced from Transition Region and Coronal Explorer Observations}",
      journal = {\apjl},
         year = 2002,
        month = sep,
       volume = {576},
       number = {2},
        pages = {L153-L156},
          doi = {10.1086/343886},
       adsurl = {https://ui.adsabs.harvard.edu/abs/2002ApJ...576L.153O}
}

@ARTICLE{2015A&A...577A...4Z,
       author = {{Zimovets}, I.~V. and {Nakariakov}, V.~M.},
        title = "{Excitation of kink oscillations of coronal loops: statistical study}",
      journal = {\aap},
         year = 2015,
        month = may,
       volume = {577},
          eid = {A4},
        pages = {A4},
          doi = {10.1051/0004-6361/201424960},
       adsurl = {https://ui.adsabs.harvard.edu/abs/2015A&A...577A...4Z}
}

@ARTICLE{2020ARA&A..58..441N,
       author = {{Nakariakov}, Valery M. and {Kolotkov}, Dmitrii Y.},
        title = "{Magnetohydrodynamic Waves in the Solar Corona}",
      journal = {\araa},
         year = 2020,
        month = aug,
       volume = {58},
        pages = {441-481},
          doi = {10.1146/annurev-astro-032320-042940},
       adsurl = {https://ui.adsabs.harvard.edu/abs/2020ARA&A..58..441N}
}

@ARTICLE{2020PPCF...62a4016A,
       author = {{Antolin}, Patrick},
        title = "{Thermal instability and non-equilibrium in solar coronal loops: from coronal rain to long-period intensity pulsations}",
      journal = {Plasma Physics and Controlled Fusion},
         year = 2020,
        month = jan,
       volume = {62},
       number = {1},
          eid = {014016},
        pages = {014016},
          doi = {10.1088/1361-6587/ab5406},
       adsurl = {https://ui.adsabs.harvard.edu/abs/2020PPCF...62a4016A}
}

@ARTICLE{2024A&A...685A..36S,
       author = {{Shrivastav}, Arpit Kumar and {Pant}, Vaibhav and {Berghmans}, David and {Zhukov}, Andrei N. and {Van Doorsselaere}, Tom and {Petrova}, Elena and {Banerjee}, Dipankar and {Lim}, Daye and {Verbeeck}, Cis},
        title = "{Statistical investigation of decayless oscillations in small-scale coronal loops observed by Solar Orbiter/EUI}",
      journal = {\aap},
         year = 2024,
        month = may,
       volume = {685},
          eid = {A36},
        pages = {A36},
          doi = {10.1051/0004-6361/202346670},
archivePrefix = {arXiv},
       eprint = {2304.13554},
 primaryClass = {astro-ph.SR},
       adsurl = {https://ui.adsabs.harvard.edu/abs/2024A&A...685A..36S}
}

@ARTICLE{2024MNRAS.531.4611N,
       author = {{Nakariakov}, Valery M. and {Zhong}, Yu and {Kolotkov}, Dmitrii Y.},
        title = "{Transition from decaying to decayless kink oscillations of solar coronal loops}",
      journal = {\mnras},
         year = 2024,
        month = jul,
       volume = {531},
       number = {4},
        pages = {4611-4618},
          doi = {10.1093/mnras/stae1483},
archivePrefix = {arXiv},
       eprint = {2406.07490},
 primaryClass = {astro-ph.SR},
       adsurl = {https://ui.adsabs.harvard.edu/abs/2024MNRAS.531.4611N}
}

@ARTICLE{1999Sci...285..862N,
       author = {{Nakariakov}, V.~M. and {Ofman}, L. and {Deluca}, E.~E. and {Roberts}, B. and {Davila}, J.~M.},
        title = "{TRACE observation of damped coronal loop oscillations: Implications for coronal heating}",
      journal = {Science},
         year = 1999,
        month = aug,
       volume = {285},
        pages = {862-864},
          doi = {10.1126/science.285.5429.862},
       adsurl = {https://ui.adsabs.harvard.edu/abs/1999Sci...285..862N}
}

@ARTICLE{2019ApJS..241...31N,
       author = {{Nechaeva}, Alena and {Zimovets}, Ivan V. and {Nakariakov}, V.~M. and {Goddard}, C.~R.},
        title = "{Catalog of Decaying Kink Oscillations of Coronal Loops in the 24th Solar Cycle}",
      journal = {\apjs},
         year = 2019,
        month = apr,
       volume = {241},
       number = {2},
          eid = {31},
        pages = {31},
          doi = {10.3847/1538-4365/ab0e86},
       adsurl = {https://ui.adsabs.harvard.edu/abs/2019ApJS..241...31N}
}

@ARTICLE{2015A&A...580A..57R,
       author = {{Ruderman}, M.~S. and {Terradas}, J.},
        title = "{Standing kink oscillations of thin twisted magnetic tubes with continuous equilibrium magnetic field}",
      journal = {\aap},
         year = 2015,
        month = aug,
       volume = {580},
          eid = {A57},
        pages = {A57},
          doi = {10.1051/0004-6361/201526168},
       adsurl = {https://ui.adsabs.harvard.edu/abs/2015A&A...580A..57R}
}

@ARTICLE{2020MNRAS.496...67B,
       author = {{Bahari}, K. and {Petrukhin}, N.~S. and {Ruderman}, M.~S.},
        title = "{Resonant damping and instability of propagating kink waves in flowing and twisted magnetic flux tubes}",
      journal = {\mnras},
         year = 2020,
        month = jul,
       volume = {496},
       number = {1},
        pages = {67-79},
          doi = {10.1093/mnras/staa1442},
       adsurl = {https://ui.adsabs.harvard.edu/abs/2020MNRAS.496...67B}
}

@ARTICLE{2024ApJ...972...38L,
       author = {{Lopin}, Igor},
        title = "{Transverse Oscillations and Kelvin Helmholtz Instability in Curved Arcade Loops with Siphon Flows}",
      journal = {\apj},
         year = 2024,
        month = sep,
       volume = {972},
       number = {1},
          eid = {38},
        pages = {38},
          doi = {10.3847/1538-4357/ad6318},
       adsurl = {https://ui.adsabs.harvard.edu/abs/2024ApJ...972...38L}
}

@ARTICLE{2008A&A...487L..17V,
       author = {{Van Doorsselaere}, T. and {Nakariakov}, V.~M. and {Young}, P.~R. and {Verwichte}, E.},
        title = "{Coronal magnetic field measurement using loop oscillations observed by Hinode/EIS}",
      journal = {\aap},
         year = 2008,
        month = aug,
       volume = {487},
       number = {2},
        pages = {L17-L20},
          doi = {10.1051/0004-6361:200810186},
       adsurl = {https://ui.adsabs.harvard.edu/abs/2008A&A...487L..17V}
}

@ARTICLE{2015ApJ...799..151G,
       author = {{Guo}, Y. and {Erd{\'e}lyi}, R. and {Srivastava}, A.~K. and {Hao}, Q. and {Cheng}, X. and {Chen}, P.~F. and {Ding}, M.~D. and {Dwivedi}, B.~N.},
        title = "{Magnetohydrodynamic Seismology of a Coronal Loop System by the First Two Modes of Standing Kink Waves}",
      journal = {\apj},
         year = 2015,
        month = feb,
       volume = {799},
       number = {2},
          eid = {151},
        pages = {151},
          doi = {10.1088/0004-637X/799/2/151},
archivePrefix = {arXiv},
       eprint = {1411.7095},
 primaryClass = {astro-ph.SR},
       adsurl = {https://ui.adsabs.harvard.edu/abs/2015ApJ...799..151G}
}

@ARTICLE{2013ApJ...777...17S,
       author = {{Srivastava}, A.~K. and {Goossens}, M.},
        title = "{X6.9-class Flare-induced Vertical Kink Oscillations in a Large-scale Plasma Curtain as Observed by the Solar Dynamics Observatory/Atmospheric Imaging Assembly}",
      journal = {\apj},
         year = 2013,
        month = nov,
       volume = {777},
       number = {1},
          eid = {17},
        pages = {17},
          doi = {10.1088/0004-637X/777/1/17},
archivePrefix = {arXiv},
       eprint = {1308.5758},
 primaryClass = {astro-ph.SR},
       adsurl = {https://ui.adsabs.harvard.edu/abs/2013ApJ...777...17S}
}

@ARTICLE{2012A&A...545A.129W,
       author = {{White}, R.~S. and {Verwichte}, E. and {Foullon}, C.},
        title = "{First observation of a transverse vertical oscillation during the formation of a hot post-flare loop}",
      journal = {\aap},
         year = 2012,
        month = sep,
       volume = {545},
          eid = {A129},
        pages = {A129},
          doi = {10.1051/0004-6361/201219856},
       adsurl = {https://ui.adsabs.harvard.edu/abs/2012A&A...545A.129W}
}

@ARTICLE{2015A&A...582A..75O,
       author = {{Ofman}, L. and {Parisi}, M. and {Srivastava}, A.~K.},
        title = "{Three-dimensional MHD modeling of vertical kink oscillations in an active region plasma curtain}",
      journal = {\aap},
         year = 2015,
        month = oct,
       volume = {582},
          eid = {A75},
        pages = {A75},
          doi = {10.1051/0004-6361/201425054},
archivePrefix = {arXiv},
       eprint = {1505.05427},
 primaryClass = {astro-ph.SR},
       adsurl = {https://ui.adsabs.harvard.edu/abs/2015A&A...582A..75O}
}

@ARTICLE{2011ApJ...726...42S,
       author = {{Selwa}, M. and {Ofman}, L. and {Solanki}, S.~K.},
        title = "{The Role of Active Region Loop Geometry. I. How Can it Affect Coronal Seismology?}",
      journal = {\apj},
         year = 2011,
        month = jan,
       volume = {726},
       number = {1},
          eid = {42},
        pages = {42},
          doi = {10.1088/0004-637X/726/1/42},
       adsurl = {https://ui.adsabs.harvard.edu/abs/2011ApJ...726...42S}
}

\end{document}